\documentclass[a4paper]{article}
\usepackage{authblk} 
\usepackage[pages=all, color=black, position={current page.south}, placement=bottom, scale=1, opacity=1, vshift=5mm]{background}
\SetBgContents{
}      

\usepackage[margin=1in]{geometry} 

\usepackage{amsmath}
\usepackage{amsthm}
\usepackage{amssymb}
\usepackage{lipsum}
\usepackage{xcolor}
\usepackage{enumerate}

\usepackage[utf8]{inputenc}
\usepackage[hidelinks]{hyperref}
\hypersetup{
	unicode,
	pdfauthor={Author One, Author Two, Author Three},
	pdftitle={A simple article template},
	pdfsubject={A simple article template},
 	pdfkeywords={article, template, simple},
	pdfproducer={LaTeX},
	pdfcreator={pdflatex}
}

\usepackage{natbib}
\theoremstyle{plain}
\newtheorem{theorem}{Theorem}[section] %
\newtheorem*{proof*}{Proof} %
\newtheorem*{theorem*}{Theorem}
\newtheorem{corollary}[theorem]{Corollary}%
\newtheorem{lemma}[theorem]{Lemma}

\newtheorem{proposition}[theorem]{Proposition} %

\theoremstyle{definition}

\newtheorem{remark}[theorem]{Remark}

\usepackage{graphicx, color}
\usepackage{booktabs}
\usepackage[table]{xcolor}
\usepackage{float}
\graphicspath{{Figures/}}

\usepackage{tikz}
\usetikzlibrary{calc}
\usetikzlibrary{positioning}
\usetikzlibrary{fit}
\usetikzlibrary{shapes,arrows}

\usepackage{algorithm, algorithmic} 
\usepackage{mathrsfs} 
\usepackage{subcaption}
\usepackage{bm}
\usepackage{xcolor}

\usepackage[overlay, absolute]{textpos}
\usepackage{xurl}
\usepackage[toc,page]{appendix}  
\usepackage{titletoc}
\usepackage[acronym]{glossaries}

\newcommand{\1}{\mathbf{1} }
\newcommand{\R}{\mathbb{R}}

\renewcommand{\P}{\mathcal{P}}

\newcommand{\norm}[1]{\lVert #1 \rVert}

\title{Robust Posterior Projection for Bayesian Inference Under Order Restrictions}

\author[1]{Imène Bouafia}
\author[1]{Anne Philippe}
\author[2]{Guillaume Guérin}

\affil[1]{Laboratoire de Mathématiques Jean Leray, UMR 6629  Nantes Université - CNRS, 44000 Nantes, France}
\affil[2]{Géosciences Rennes, UMR 6118 Univ Rennes - CNRS, 35000 Rennes, France}

\date{}

\begin{document}
\maketitle

\begin{abstract}

  In archaeology and geochronology, stratigraphic information constrains the relative chronological order of dated samples. Bayesian chronological models incorporate these constraints either through the prior, or by projecting an unconstrained posterior distribution onto the set of ordered chronologies. Dating measurements may nevertheless appear inconsistent with their stratigraphic position, owing to disturbances that do not justify discarding them. We introduce  different robust projection maps based on the Huber loss and characterize the non-uniqueness of the considered maps which justifies a Tikhonov regularization. The robust maps also decompose each sample into a monotone baseline and a spike, which flags questionable samples. We adapt the Sequential Block Merging algorithm for the Huber-type losses, and propose a data-driven heuristic for selecting the  hyperparameter on which each loss depends on. Simulation studies and geochronological case studies illustrate the gains in robustness and their limits.

\end{abstract}

\tableofcontents
\clearpage

\section{Introduction}
Constructing numerical chronologies in the fields of archaeology and geomorphology often relies on stratigraphic
information to constrain the relative ordering of dated samples. In
fact, during excavation and sampling, samples collected from different stratigraphic
units are dated using methods such as $ ^{14}C$ or Luminescence for the last tens to hundreds ka or so. To
reduce uncertainties and refine chronological inferences, it is frequent to use statistical modeling and
incorporate spatial information: according to the principle of stratigraphy, the age $\theta_j$ of a given
sample is greater than the ages of all samples lying above it. These inequalities are
called stratigraphic constraints and are incorporated as prior knowledge in Bayesian
chronological models such as Chronomodel \citep{lanos_event_2018}, BayLum \citep{combes_bayesian_2017} or OxCal \citep{bronk_ramsey_bayesian_2009}.  
\citet{bouafia2026projectionbasedapproachbayesianage}  proposed a projected-posterior approach as a way of introducing the stratigraphic constraints in the Bayesian modeling. Rather than modifying the prior distribution to enforce the order, the latter approach first estimates the posterior distribution from the dating measurements alone and subsequently projects posterior samples onto the space of chronologies satisfying the specified constraints. This projection can be formulated as a weighted isotonic regression problem. It provides a minimally informative way of incorporating stratigraphic information while preserving the variability supported by the measurements. 

\noindent The posterior projection, however, assumes that each measurement is informative about the age of the stratigraphic unit from which the sample was taken. In practice, this assumption may fail at the level of individual samples. For example, a
sample may be intrusive because of post-depositional processes (e.g., bioturbation), so that the dated material  does not belong to the depositional event of its presumed unit; or the luminescence signal emitted by a sample may not have been fully reset at the time of deposition, leading to age overestimation; or a radiocarbon sample may have incorporated modern (or old) carbon. In each case, the reported age carries a systematic
offset with respect to the age of its stratigraphic position, while the
stratigraphic relations themselves, directly inferred from sampled sections and their relationships, remain valid.

\noindent The treatment of observations that are inconsistent with the ordering constraints, often called stratigraphic inversions, has already been considered in Bayesian chronological modeling. \citet{christen1994summarizing} introduced an explicit probabilistic outlier formulation, in which each dated sample may be associated with a latent offset from the chronological quantity of interest. An outlier indicator determines whether this additional discrepancy is active, while the offset itself is treated as an unobserved quantity rather than being imposed by the chronological constraints. \citet{BronkRamseyOutlierss_2009} subsequently developed this framework in the context of radiocarbon modeling with OxCal.  These models therefore allow observations that are poorly compatible with the chronological model to be accommodated through an alternative outlier component, rather than requiring them to be removed a priori.

\noindent Alternatively, a projection onto the isotonic cone treats such observations as exact. Consequently, a single sample whose age is displaced relative to its unit can exert a disproportionate influence on the projected chronology, including on the estimates of neighboring samples whose measurements are unproblematic.\\
This often-encountered issue raises the question of how stratigraphic information can be incorporated
when a small number of measurements are incompatible with the assumed
ordering. In particular, we ask whether the projected-posterior approach can
be extended to accommodate such observations while preventing them from
substantially distorting the chronological inferences. Errors in the
stratigraphic record itself, such as missing or reversed relations in the
directed graph corresponding to stratigraphic constraints, raise a different question; thus, they are not considered in the present work.

\noindent The main contributions of this paper are threefold. First, we extend the framework of posterior projection in geochronological applications of \citet{bouafia2026projectionbasedapproachbayesianage} to robust projection maps, allowing for the treatment of potentially contaminated measurements in the presence of stratigraphic constraints. We introduce three robust projection maps based on alternative loss functions and establish theoretical properties of the resulting projected posteriors, including conditions under which they approximate the standard quadratic posterior projection.

\noindent Second, we develop computational strategies for implementing these robust projections. In particular, we adapt the Sequential Block Merging (SBM) algorithm of \citet{wang_efficient_2022} to the Huber loss and related robust loss functions. This extension also provides numerical results for SBM beyond the quadratic setting considered in the original paper.

\noindent Third, we demonstrate the practical relevance of the proposed methodology through four geochronological case studies, illustrating its application to chronological inference in the presence of potentially contaminated measurements.

\noindent The remainder of the paper is organized as follows. Section~\ref{sec:background} reviews Bayesian chronological inference under stratigraphic constraints, with a particular focus on posterior projection methods and their limitations in the presence of substantial measurement offsets. Section~\ref{sec:projection_maps} introduces three robust projection maps designed to accommodate such offsets. We formulate the corresponding isotonic projection problems and introduce a Tikhonov regularization to ensure uniqueness, when necessary. We also establish theoretical guarantees relating the robust projections to the standard quadratic isotonic projection when the hyperparameters exceed appropriate thresholds. Section~\ref{sec:computational} presents the computational strategies used to solve the robust projection problems through the adaptation of the Sequential Block Merging algorithm. Section~\ref{sec:hyperparam} establishes the three robust posterior projections and introduces a practical pipeline for selecting the unique hyperparameter for each. Section~\ref{sec:simulation} illustrates the resulting projections through four simulation studies. Finally, Section~\ref{sec:real_applications} presents four geochronological case studies illustrating the proposed methodology.


\section{Background} \label{sec:background}

Let $M_1,\ldots,M_n$ denote the age measurements obtained from $n$ dated
samples. We consider these measurements as observations from a parametric family distribution  $\lbrace f(. \mid \theta), \ \theta \in  \R^n \rbrace$. 
Stratigraphic information is represented by a directed acyclic graph (DAG)
$G=(V,E)$, where $V=\{1,\ldots,n\}$ represents the dated samples and each
directed edge $(i,j)\in E$ encodes an observed stratigraphic relationship,
corresponding to the chronological constraint
$\theta_i\leq\theta_j$. The set of parameter values satisfying all stratigraphic constraints is therefore the isotonic cone
\begin{equation*}
    \mathcal{P}
    =
    \left\{
        \theta\in\mathbb{R}^n
        \mid
        \theta_i\leq\theta_j,\quad \forall (i,j)\in E
    \right\}.
  \end{equation*}
Every edge $(i,j) \in E$ encodes a stratigraphic relation, so that
the true chronology $\theta$ belongs to $\mathcal{P}$. Robustness is sought with
respect to individual measurements that are incompatible with $G$, not with
respect to errors in $G$ itself.
  
\noindent The following Bayesian approaches can be used to incorporate the stratigraphic constraints.
For a total ordering,
$\mathcal{P}=\{\theta\in\mathbb{R}^n:
\theta_1\leq\cdots\leq\theta_n\}$, we consider the classical Uniform-order model, it
incorporates the ordering directly through the prior
\begin{equation}
    \label{eq:uodensity}
    \pi_{\mathcal{P}}(\theta)
   \propto
    \mathbf{1}_{\mathcal{P}\cap[\alpha,\beta]^n}(\theta),
\end{equation}
where $[\alpha,\beta]$ denotes the assumed study period. The corresponding
posterior is
\begin{equation*}
    \pi_{\mathcal{P}}(\theta\mid M)
    \propto
    f(M\mid\theta)\,
    \pi_{\mathcal{P}}(\theta).
\end{equation*}
This formulation incorporates the stratigraphic order directly into the
Bayesian model and consequently restricts the posterior distribution to the
constrained space.
\noindent As an alternative, the measurements can first be analyzed without
stratigraphic constraints. We refer to this as the Unconstrained model.
It can be defined using either a flat prior,
\begin{equation}
    \label{eq:flat}
    \pi(\theta)\propto 1_{\R^n},
\end{equation}
or a uniform prior over the study period,
\begin{equation}
    \label{eq:uniform_prior}
    \pi(\theta)
    =
    \frac{\mathbf{1}_{[\alpha,\beta]^n}(\theta)}
    {(\beta-\alpha)^n}.
\end{equation}
In both cases, the resulting posterior is
\begin{equation} \label{eq:unconstrained}
    \pi(\theta\mid M)
    \propto
    f(M\mid\theta)\,\pi(\theta).
\end{equation}
The projected-posterior approach introduced in \citet{bouafia2026projectionbasedapproachbayesianage} provides a
different way of incorporating stratigraphic information. Rather than
including the ordering constraints in the prior, posterior samples are first
drawn from the Unconstrained posterior and subsequently projected onto
$\mathcal{P}$. Let $\Pi(\cdot\mid M)$ denote the Unconstrained posterior
measure. The projected posterior is defined as the push-forward measure
\begin{equation}
    \label{eq:pushforward}
    \widetilde{\Pi}(\cdot\mid M)
    =
    (T_{\mathcal{P}})_{\#}
    \Pi(\cdot\mid M),
\end{equation}
where $T_{\mathcal{P}}$ denotes the projection onto the isotonic cone, which must be well-defined i.e. for any $\theta' \in \R^n$, its projection $T_\P(\theta')$ is unique.

\noindent \cite{bouafia2026projectionbasedapproachbayesianage} consider the weighted Euclidean projection
\begin{equation}
    \label{eq:IR}
    T_{\mathcal{P}}(\theta')
    =
    \underset{\theta\in\mathcal{P}}{\arg\min}
    \left\|
        W^{1/2}(\theta-\theta')
    \right\|_2^2 := \arg \min_{\theta \in \P} Q(\theta),
\end{equation}
where \begin{equation} \label{eq:W}
  W = \textrm{diag}(w_1, \cdots, w_n), 
\end{equation}
is a diagonal matrix and the weights are chosen to be \begin{equation*}
    w_i
    =
    \operatorname{Var}(\theta_i\mid M)^{-1},
    \qquad i=1,\ldots,n,
\end{equation*}
the inverse posterior variances under the Unconstrained model.
The objective function can be written as a separable function, i.e
\begin{equation} \label{eq:quadratic_objective}
    \sum_{i=1}^n w_i(\theta_i-\theta_i')^2.
\end{equation}

\noindent The resulting optimization problem is an isotonic regression problem on the
DAG.
In \cite{bouafia2026projectionbasedapproachbayesianage}, the projected posterior has been shown to provide a useful way of
incorporating stratigraphic information while preserving the variability of
the Unconstrained posterior. In particular, it can substantially reduce the
duration inflation associated with classical order-constrained Bayesian
models when true ages are closely spaced relative to measurement
uncertainties. However, the projected posterior remains dependent on the
quality of the Unconstrained posterior and, consequently, on the behavior of
the projection when individual observations strongly contradict the assumed
stratigraphic order.

\noindent We refer to such observations as \emph{stratigraphic-order outliers}. A contaminated measurement
$M_i$ is indistinguishable, on the basis of that measurement alone, from an
uncontaminated measurement of an older or younger sample: no marginal
diagnostic applied to $M_i$ can flag it. It is anomalous only relative to the
ordering encoded by $G$. Robustness can therefore not be introduced at the
level of the Unconstrained posterior, which never sees the constraints; it
must be built into the projection step. Since both the Uniform-order model
and the projected-posterior approach enforce the stratigraphic constraints, a
strongly inconsistent observation may affect not only its own estimate but
also the estimates of neighboring samples.

\begin{figure}
  \centering
  \includegraphics[width=\linewidth]{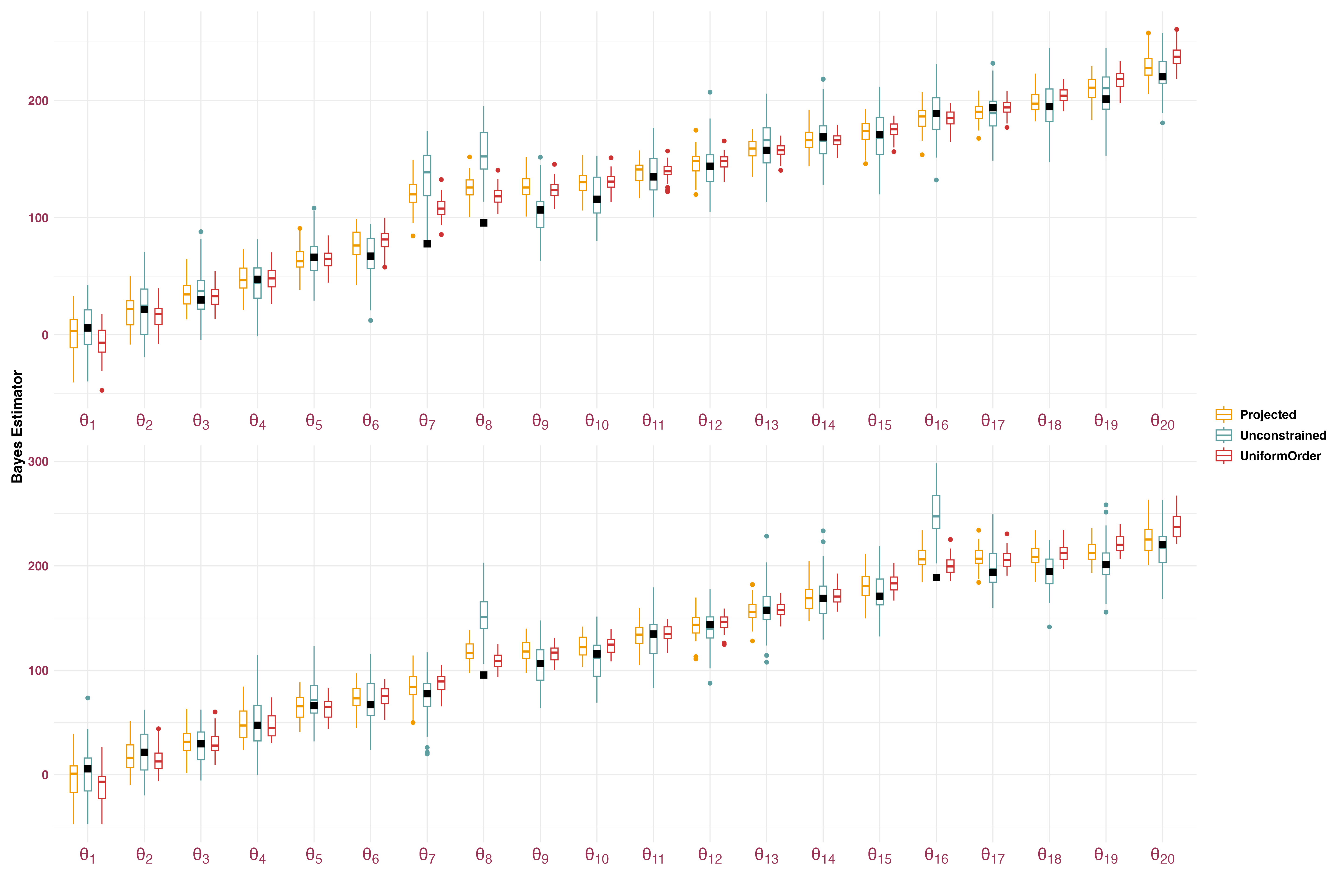}
  \caption{Boxplots of Bayes estimates obtained from 50 replications of
  the simulated measurements under the contamination scenario
  \eqref{eq:contamination_scenario}. Here, $a=(60,60)$,
  $\sigma=20$ for all $i$, with $q_a=\{7,8\}$ (top) and
  $q_a=\{8,16\}$ (bottom).}
  \label{fig:contamination_scenario}
\end{figure}

\noindent To investigate this behavior, we consider a controlled contamination
scenario in which the true chronology is perfectly ordered. We assume the
measurement model
\begin{equation}
    \label{eq:simulation_model}
    M_i
    \sim
    \mathcal{N}(\theta_i,\sigma_i^2),
    \qquad i=1,\ldots,n,
\end{equation}
where
$\theta=(\theta_1,\ldots,\theta_n)$ represents the true ages satisfying
\[
    \theta_1\leq\cdots\leq\theta_n,
\]
and $\sigma=(\sigma_1,\ldots,\sigma_n)$ contains the known measurement
uncertainties.
We then select a subset of samples
\begin{equation*}
    q_a=\{i_1,\ldots,i_k\} \subset \{1,\ldots,n \}\qquad  k  \ll n,
\end{equation*}
and introduce a systematic deviation in their measurements. For each
contaminated sample, we consider
\begin{equation}
    \label{eq:contamination_scenario}
    M_{i}
    \sim
    \mathcal{N}
    \left(
        \theta_{i}+a_{i},
        \sigma_{i}^2
    \right),
    \qquad i\in q_a,
\end{equation}
where $(a_{i})_{i\in q_a }$ denotes the contamination magnitude. The remaining
measurements  $M_i, \; i\notin q_a$ follow the uncontaminated model in
Equation~\eqref{eq:simulation_model}. Figure~\ref{fig:contamination_scenario} illustrates the effect of this
contamination on the different Bayesian approaches. Both the Uniform-order
and projected-posterior models can be substantially affected by the presence
of stratigraphic-order outliers. In particular, a contaminated observation
can induce under or over-estimation of neighboring uncontaminated samples. The resulting
effect is a consequence of the way in which the ordering constraint is
enforced: a strongly deviating observation cannot be discarded by the
projection and must instead be reconciled with the remaining observations
while satisfying the order constraints.

\noindent The mechanism underlying this behavior can be seen directly from the
isotonic regression problem. 
The Karush-Kuhn-Tucker (KKT) conditions allow one to write the optimal solution into partitioned blocks

\[
    V =    B_1 \cup \ldots \cup B_K \quad K \in \{1, \cdots, n\},
\]
where each block $B_o, \; o \in \{1, \cdots, K\}$, is a set of samples $i \in V$ and within $B_o$, the optimal solution is constant obtained from the scalar optimization problem
\begin{equation*}
    \theta_o
    =
    \underset{t\in\mathbb{R}}{\arg\min}
    \sum_{i\in B_o}
    w_i(\theta_i'-t)^2.
\end{equation*}
For the quadratic objective function \eqref{eq:quadratic_objective}, the unique minimizer is the weighted mean
\begin{equation}\label{eq:block_mean}
    \theta_o
    =
    \frac{
        \sum_{i\in B_o}w_i\theta_i'
    }{
        \sum_{i\in B_o}w_i
    }.
\end{equation}

\noindent Consequently, a contaminated observation with a positive or negative
deviation can shift the common value of its block in the corresponding
direction. This shift also affects uncontaminated observations belonging to
the same block. Moreover, when the deviation is sufficiently large, the
violation of the ordering constraints can lead to the formation of larger
blocks, thereby propagating the effect of the contaminated observation to
additional neighboring samples. This behavior is illustrated, for example,
by the estimates of samples 9 and 10 in the upper panel of
Figure~\ref{fig:contamination_scenario}.

\begin{proposition}\label{prop:drift-quadratic}
Let $\mathcal P=\{\theta_1\le\dots\le\theta_n\}$,  $e_i$, the $i$-th canonical basis vector of $\mathbb R^n$ and  $T_{\mathcal P}$, the weighted projection~\eqref{eq:IR}. For a data vector $m\in\mathbb R^n$,
$\Pi(\cdot\,;m)$  is the unconstrained posterior under the flat prior
\eqref{eq:flat} and the Gaussian likelihood of
\eqref{eq:simulation_model}. Consider $m+ae_i$ the data $m$ with the $i$-th
measurement contaminated as in \eqref{eq:contamination_scenario} with $q_a=\{i\}$ and
$a_i=a$. Set $\Pi_a=\Pi(\cdot\,;m+ae_i)$ and $\tilde\Pi_a=(T_{\mathcal P})_\#\Pi_a$.
Then the projected posterior mean of the $i$-th age,
\[
\bar\theta_i(a)=\int_{\mathbb R^n}\vartheta_i\,\tilde\Pi_a(d\vartheta),
\]
exists and satisfies
\[
\frac{w_i}{W_n}\,a\;\le\;\bar\theta_i(a)-\bar\theta_i(0)\;\le\;a,
\qquad\forall a\ge0, 
\]
where $W_n=\sum_{j=1}^n w_j$. 
\end{proposition}
\begin{proof}
For the total order $\mathcal P$, the max--min
characterization of isotonic regression gives \citep[see, e.g.,][]{robertson_order_1988}
\[
T_{\mathcal P}(z)_i=\max_{s\le i}\ \min_{t\ge i}\ A_{s:t}(z),
\qquad A_{s:t}(z)=\frac{\sum_{j=s}^t w_jz_j}{\sum_{j=s}^t w_j}.
\]
Since $s\le i\le t$, every interval $\{s,\dots,t\}$ in this formula contains $i$, so,
for all $\theta'\in\mathbb R^n$,
\[
A_{s:t}(\theta'+ae_i)=A_{s:t}(\theta')+\frac{w_i}{\sum_{j=s}^t w_j}\,a .
\]
As $w_i/W_n\le w_i/\sum_{j=s}^t w_j\le 1$ and $a\ge0$, we get
\[
A_{s:t}(\theta')+\frac{w_i}{W_n}\,a\;\le\;A_{s:t}(\theta'+ae_i)\;\le\;A_{s:t}(\theta')+a .
\]
Both $\max$ and $\min$ are non-decreasing in each argument and
translation-equivariant. Applying this first to the inner minimum over $t$, then to
the outer maximum over $s$, yields 
\begin{equation}\label{eq:bound_contam}
\frac{w_i}{W_n}\,a\;\le\;T_{\mathcal P}(\theta'+ae_i)_i-T_{\mathcal P}(\theta')_i\;\le\;a,
\qquad\forall\theta'\in\mathbb R^n .
\end{equation}
The distribution $\Pi_a=\Pi(\cdot\,;m+ae_i)$ is the gaussian distributions with mean $m+ae_i$ and covariance matrix $\operatorname{diag}(\sigma_1^2, \cdots, \sigma_,^2)$. Denoting by
$\pi(\cdot\,;m)$ the density of $\Pi(\cdot\,;m)$, we have
$\pi(x\,;m+ae_i)=\pi(x-ae_i\,;m)$ for all $x\in\mathbb R^n$. By~\eqref{eq:block_mean},
$T_{\mathcal P}(z)_i$ is a convex combination of the coordinates of $z$, so $|T_{\mathcal P}(z)_i|\le\max_j|z_j|\le\sum_{j=1}^n|z_j|$.  The Bayes estimate $\bar\theta_i(a)$ exists since $\Pi_a$ is Gaussian. By definition of the push-forward and the change of variables
$x\mapsto x+ae_i$,
\[
\bar\theta_i(a)=\int_{\mathbb R^n}T_{\mathcal P}(x)_i\,\pi(x\,;m+ae_i)\,dx
=\int_{\mathbb R^n}T_{\mathcal P}(x+ae_i)_i\,\pi(x\,;m)\,dx .
\]
Hence
\begin{equation}\label{eq:diff_mean}
\bar\theta_i(a)-\bar\theta_i(0)
=\int_{\mathbb R^n}\bigl[T_{\mathcal P}(x+ae_i)_i-T_{\mathcal P}(x)_i\bigr]\,
\pi(x\,;m)\,dx ,
\end{equation}
and the result follows by integrating~\eqref{eq:bound_contam}. 
\end{proof}
\noindent Proposition~\ref{prop:drift-quadratic} shows that the quadratic projection does not
protect the posterior mean of a contaminated sample: under a contamination $a$, its
projected posterior mean is biased by at least $a\,w_i/W_n$, linearly in
$a$. The upper bound shows that the projected mean never moves by more than the contamination itself.\\
The bounds \eqref{eq:bound_contam} hold for each posterior sample $\theta'$, so they transfer to any quantile:
writing $q_\alpha(\cdot)$ for the $\alpha$-quantile of the $i$-th marginal,
\[
q_\alpha\bigl(\tilde\Pi_0\bigr)+\frac{w_i}{W_n}\,a\;\le\;q_\alpha\bigl(\tilde\Pi_a\bigr)
\;\le\;q_\alpha\bigl(\tilde\Pi_0\bigr)+a .
\]
The whole credible interval of a contaminated sample is therefore shifted in the
direction of the contamination, which explains the loss of coverage of the
quadratic projections at contaminated samples (see Table \ref{tab:coverage_posM80}).

\noindent These results motivate the use of robust loss functions. More generally,
we seek an objective function that reduces the influence of observations
whose deviations are incompatible with the assumed stratigraphic structure.
This can be achieved by reducing the contribution of large residuals to the
optimization problem, thereby limiting their ability to modify the estimates
of otherwise consistent observations. The remainder of this work investigates
such robust formulations for the projection step.

\section{Alternative losses for the projection map} \label{sec:projection_maps}

\subsection{Huber Loss} 

The Huber loss \citep{huber_robust_1964} is a piecewise, convex function, quadratic for small residuals $r$ and linear for large ones: \begin{equation}
  \label{eq:huber_loss}
	\rho_\delta(r) = \begin{cases}
		r^2 &\text{if } \mid r\mid \leq \delta \\
		2\delta \mid r \mid - \delta^2 &\text{else}
	\end{cases},
\end{equation}
where $\delta > 0$ is a hyperparameter to be fixed. 
\noindent The projection map reads:
\begin{equation} \label{eq:huber_map}
	T_\P^{(\delta)} (\theta') = \arg\min_{\theta \in \P}  \sum_{i=1}^{n} w_i \rho_\delta(\theta'_i-\theta_i) =:  \arg\min_{\theta \in \P} H_{\delta}(\theta) \quad \forall \ \theta' \in \R^n. 
\end{equation}

\begin{proposition}\label{prop:existence-huber}
The objective function $H_\delta$ is convex and coercive hence it admits at least one minimizer. The map $T_\P^{(\delta)}$ is therefore set-valued,
and for every $\theta' \in \R^n$  the set $T_\P^{(\delta)}(\theta')$  is non-empty, closed and convex.
\end{proposition}
\begin{proof}
The function $H_{\delta}$ is continuous, and coercive.
We denote $$\rho^* = \min_{\theta \in \P}  H_{\delta}(\theta),$$ the
minimum value of $H_{\delta}$. Therefore, by continuity of the
objective function,  $$T_\P^{(\delta)}(\theta') =
H_{\delta}^{-1}(\lbrace \rho^* \rbrace) \cap \P$$ is closed.
Since the objective function is convex and the feasible set
$\mathcal P$ is convex, the set of minimizers $T_{\mathcal
  P}^{(\delta)}(\theta')$ is also convex. Indeed, if $\theta_1,
\theta_2 \in T_\P^{(\delta)}(\theta')$, then for all $ t \in \lbrack
0,1 \rbrack$: $t\theta_1 + (1-t)\theta_2 \in \P$\begin{align*}
	H_{\delta}( \theta_1) = \rho^* \quad \text{ and } \quad H_{\delta}(\theta_2) = \rho^* \\
	H_{\delta}(t\theta_1 + (1-t)\theta_2) \underset{\text{convexity}}{\leq} tH_{\delta} (\theta_1) + (1-t) H_{\delta}( \theta_2) = \rho^*
\end{align*}
Hence, $t\theta_1 + (1-t)\theta_2 \in T_\P^{(\delta)}(\theta')$. 
\end{proof}

\noindent By convention, we set $\min\left( 1; \dfrac{1}{|x|}\right) = 1$ if $x = 0$.  
\begin{proposition} \label{prop:existence}
Let $B_1 \cup \cdots \cup B_K$ be the partition of an optimal solution from $T_\P^{(\delta)}(\theta')$. Over each block $B_o, \; o \in \{1, \cdots, K\}$, the KKT stationarity conditions reduce to the scalar equation
\[
h_\delta(\theta_o) = 0,
\]
where
\[
h_\delta(\theta) = -2 \sum_{i \in B_o} w_i \min\!\left(1, \frac{\delta}{|\theta'_i - \theta|} \right)(\theta'_i - \theta) \quad  \forall \theta \in \R.
\]
The equation admits at least one solution.
\end{proposition}

\begin{proof}
The function $h_\delta$ is continuous, non-decreasing and satisfies
\[
\lim_{\theta \to -\infty} h_\delta(\theta) = -2\delta \sum_{i \in B_o} w_i,
\quad
\lim_{\theta \to +\infty} h_\delta(\theta) = 2\delta \sum_{i \in B_o} w_i.
\]
The result follows from the intermediate value theorem.
\end{proof}
\noindent Under the projection map $T_\P^{(\delta)}$, defined in \eqref{eq:huber_map}, we can take a closer look at the solution over blocks, and see how the non-uniqueness affects it. 
\begin{proposition} \label{prop:huber_unic}
Let $\delta>0$, let $(\theta'_{(1)},\dots,\theta'_{(p)})$ be the ordered values of $(\theta'_i)_{i\in B_o}$,
with associated positive weights $(w_{(1)},\dots,w_{(p)})$, and define
\[
W_k=\sum_{j=1}^{k}w_{(j)},\qquad W_o=\sum_{i\in B_o}w_i .
\]
Then the solution set of $h_\delta(\theta)=0$ contains a non-trivial interval if and only if there exists
$k\in\{1,\dots,p-1\}$ such that
\[
\theta'_{(k+1)}-\theta'_{(k)}>2\delta
\qquad\text{and}\qquad
W_k=\tfrac12 W_o .
\]
\end{proposition}

\begin{proof}
Assume such a $k$ exists. For any $\theta\in[\theta'_{(k)}+\delta,\ \theta'_{(k+1)}-\delta]$, which is a
non-trivial interval since $\theta'_{(k+1)}-\theta'_{(k)}>2\delta$, we have
\[
\theta'_{(j)}-\theta\le-\delta \ \text{ for } j\le k,
\qquad
\theta'_{(j)}-\theta\ge\delta \ \text{ for } j\ge k+1 .
\]
Thus
\[
h_\delta(\theta)=-2\delta\Bigl(-\sum_{j=1}^{k}w_{(j)}+\sum_{j=k+1}^{p}w_{(j)}\Bigr)
=-2\delta\,(W_o-2W_k)=0 ,
\]
and every $\theta$ in this interval is a solution.

\noindent Conversely, assume there exist $c_1<c_2$ such that $h_\delta\equiv0$ on $[c_1,c_2]$. Write
$h_\delta(c)=\sum_{i\in B_o}\varphi_i(c)$ with
$\varphi_i(c)=2w_i\max\bigl(-\delta,\min(\delta,\,c-\theta'_i)\bigr)$.
Each $\varphi_i$ is non-decreasing and their sum is constant on $[c_1,c_2]$, so each $\varphi_i$ is
constant on $[c_1,c_2]$. Since $\varphi_i$ is strictly increasing on $[\theta'_i-\delta,\theta'_i+\delta]$,
this forces $\theta'_i\le c_1-\delta$ or $\theta'_i\ge c_2+\delta$. Let
$k=\#\{i\in B_o:\theta'_i\le c_1-\delta\}$; as the two groups are separated, these are the $k$ smallest
values, so that
\[
\theta'_{(k)}\le c_1-\delta<c_2+\delta\le\theta'_{(k+1)}
\qquad\text{and}\qquad
0=h_\delta(c_1)=2\delta W_k-2\delta(W_o-W_k).
\]
Hence $W_k=\tfrac12W_o$, which excludes $k\in\{0,p\}$ since $W_o>0$, and
$\theta'_{(k+1)}-\theta'_{(k)}\ge c_2-c_1+2\delta>2\delta$.

\end{proof}
\noindent Since $h_\delta$ is continuous, non-decreasing, and satisfies $h_\delta(\theta)<0$ for $\theta$ small
enough and $h_\delta(\theta)>0$ for $\theta$ large enough (Proposition~\ref{prop:existence}), its zero
set is a non-empty closed interval. Uniqueness of the solution is therefore equivalent to this interval
being reduced to a single point, which yields the following corollary.
\begin{corollary}\label{cor:unique-root}
The equation $h_\delta(\theta)=0$ admits a unique solution if and only if, for all
$k\in\{1,\dots,p-1\}$,
\[
\theta'_{(k+1)}-\theta'_{(k)}\le2\delta
\qquad\text{or}\qquad
W_k\neq\tfrac12W_o .
\]
\end{corollary}

\begin{figure}[H]
	\centering
	\includegraphics[scale = .4]{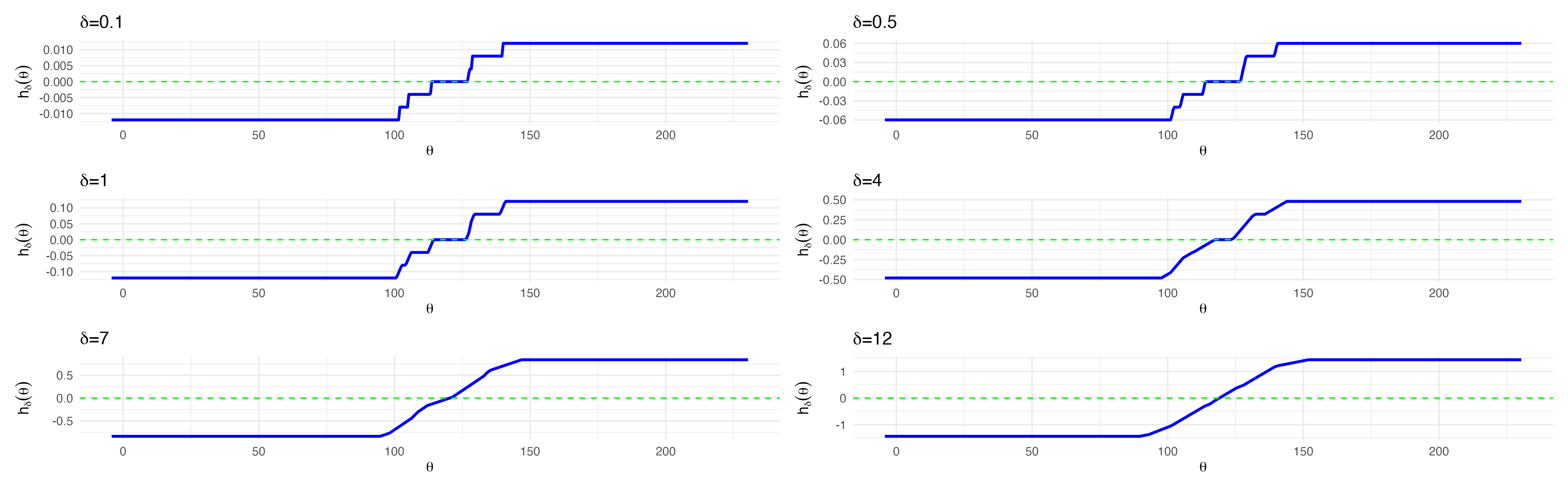}
	\caption{Example of several $h_\delta$ functions for a block of length $\lvert B_o\rvert = 6$, with equal weights $w_i = 0.1$, for which the weight balance and gap conditions are satisfied. The split index is $k = 3$, and the green dashed line indicates $0$. }
  \label{fig:h_delta}
\end{figure}

Non-uniqueness occurs only if two independent conditions hold:
\begin{itemize}
\item a gap condition: $\theta'_{(k+1)} - \theta'_{(k)} > 2\delta$,
\item a weight balance condition: $W_k = \tfrac{1}{2} W_o$.
\end{itemize}
Figure~\ref{fig:h_delta} illustrates how the gap condition holds for small values of \(\delta\) but fails as \(\delta\) increases, while the weight balance condition remains satisfied.

\noindent Since the posterior projection requires \(T_{\mathcal P}\) to return a unique solution, uniqueness must be ensured for each block. However, enforcing uniqueness by imposing additional structure on the values \(\theta' \in \R^n\) and their associated weights \(W\), as presented above,  would introduce assumptions that are not justified in the present application. We therefore seek an alternative mechanism for ensuring uniqueness without imposing such restrictions.

\subsection{Tikhonov regularization}
Uniqueness can be enforced by adding  $\varepsilon
\mid\mid \theta -c\mid \mid_W^2$ to the objective where $c\in \P$ and
$\varepsilon >0$ are fixed. This follows the classical Tikhonov regularization principle (see for instance \cite{engl1996regularization}). The resulting objective function becomes strongly convex.

\begin{proposition}
The regularized map:
	\begin{equation} \label{eq:huber_reg_map}
	\begin{aligned}
		T_\P^{(\delta, \varepsilon)}(\theta') = \arg\min_{\theta
                  \in \P} H_\delta(\theta) + \varepsilon \mid \mid \theta
                - c \mid \mid_W^2,
	\end{aligned}
\end{equation}
has a strictly convex and coercive objective, thus $T_\P^{(\delta, \varepsilon)}(\theta')$ admits a unique solution $ \theta_\varepsilon$. \\
If the Huber loss admits a unique solution, i.e $T_\P^{(\delta)}(\theta') =
\lbrace \theta^* \rbrace$ then
$$\theta_\varepsilon \underset{\varepsilon \rightarrow 0}{\longrightarrow} \theta^*.$$
If $T_\P^{(\delta)}(\theta')$ is not reduced to a
singleton, then the regularized minimizer converges to the minimizer closest to $c$ in $\norm{\cdot}_W$ : \begin{equation} \label{eq:tikhs_minimizer}
	\theta_\varepsilon \underset{\varepsilon \rightarrow 0}{\longrightarrow} \underset{\theta \in T_\P^{(\delta)}(\theta')}{\arg \min} \norm{\theta-c}_W^2
\end{equation}
\end{proposition}
\begin{proof}
For a fixed $\theta'\in\mathbb R^n$, set
$H_{(\varepsilon,\delta)}(\theta)=H_\delta(\theta)+\varepsilon\|\theta-c\|_W^2$.
Since $H_\delta$ is convex and $\theta\mapsto\varepsilon\|\theta-c\|_W^2$ is strongly convex,
$H_{(\varepsilon,\delta)}$ is strongly convex, hence coercive, and admits a unique minimizer
$\theta_\varepsilon$ over the closed convex set $\mathcal P$.

\noindent Let $\theta^*\in T^{(\delta)}_{\mathcal P}(\theta')$. Since $\theta_\varepsilon$ minimizes
$H_{(\varepsilon,\delta)}$ over $\mathcal P$ and $\theta^*$ minimizes $H_\delta$ over $\mathcal P$, we have
\begin{align*}
H_{(\varepsilon,\delta)}(\theta_\varepsilon)&\le H_{(\varepsilon,\delta)}(\theta^*),\\
H_\delta(\theta^*)&\le H_\delta(\theta_\varepsilon).
\end{align*}
Combining both inequalities yields
\begin{equation}\label{eq:bounded}
\|\theta_\varepsilon-c\|_W^2\le\|\theta^*-c\|_W^2
\qquad\text{for all }\theta^*\in T^{(\delta)}_{\mathcal P}(\theta'),
\end{equation}
and
\[
\min_{\theta\in\mathcal P}H_\delta(\theta)\le H_\delta(\theta_\varepsilon)
\le H_\delta(\theta^*)+\varepsilon\|\theta^*-c\|_W^2 ,
\]
where the left inequality holds because $\theta_\varepsilon\in\mathcal P$. Hence
\[
\lim_{\varepsilon\to0}H_\delta(\theta_\varepsilon)=\min_{\theta\in\mathcal P}H_\delta(\theta).
\]

\noindent By \eqref{eq:bounded}, $(\theta_\varepsilon)_{\varepsilon>0}$ is bounded. Let $(\varepsilon_k)$ be any
sequence with $\varepsilon_k\to0$ such that $(\theta_{\varepsilon_k})$ converges, and denote its limit by
$\bar\theta$; such sequences exist by boundedness. Since $\theta_{\varepsilon_k}\in\mathcal P$ for all $k$
and $\mathcal P$ is closed, $\bar\theta\in\mathcal P$. By continuity of $H_\delta$,
\[
H_\delta(\bar\theta)=\lim_{k\to+\infty}H_\delta(\theta_{\varepsilon_k})=\min_{\theta\in\mathcal P}H_\delta(\theta),
\]
hence $\bar\theta\in T^{(\delta)}_{\mathcal P}(\theta')$.

\noindent By Proposition~\ref{prop:existence-huber}, $T^{(\delta)}_{\mathcal P}(\theta')$ is non-empty, closed and
convex. Since $x\mapsto\|x-c\|_W^2$ is strictly convex and coercive, it admits a unique minimizer over
this set,
\[
\tilde\theta=\arg\min_{\theta\in T^{(\delta)}_{\mathcal P}(\theta')}\|\theta-c\|_W^2 .
\]
As $\bar\theta\in T^{(\delta)}_{\mathcal P}(\theta')$, we have $\|\tilde\theta-c\|_W^2\le\|\bar\theta-c\|_W^2$.
On the other hand, applying \eqref{eq:bounded} with $\theta^*=\tilde\theta$ along $(\varepsilon_k)$ and
letting $k\to+\infty$ gives $\|\bar\theta-c\|_W^2\le\|\tilde\theta-c\|_W^2$. Thus $\bar\theta$ also
minimizes $\|\cdot-c\|_W^2$ over $T^{(\delta)}_{\mathcal P}(\theta')$, and by uniqueness
$\bar\theta=\tilde\theta$.

\noindent Therefore every convergent subsequence $(\theta_{\varepsilon_k})$ with $\varepsilon_k\to0$ has limit
$\tilde\theta$. If $\theta_\varepsilon\not\to\tilde\theta$, there would exist $\eta>0$ and
$\varepsilon_k\to0$ with $\|\theta_{\varepsilon_k}-\tilde\theta\|_W\ge\eta$; by boundedness,
$(\theta_{\varepsilon_k})$ admits a convergent subsequence, whose limit is $\tilde\theta$, a
contradiction. Therefore $\theta_\varepsilon\to\tilde\theta$ as $\varepsilon\to0$.
\end{proof}

\noindent The Tikhonov regularization allows us to enforce uniqueness
by selecting a unique solution from the set of all solutions of the
isotonic regression problem under the Huber loss \(T_{\mathcal
  P}^{(\delta)}(\theta')\).  As $\varepsilon\to0$, the regularized solution converges to the Huber
solution closest to $c$.

\noindent  In the limit $\varepsilon\to0$, the center acts only as a
selection rule among the 
minimizers of the Huber problem. For a fixed $\varepsilon>0$, however, the center always
influences the solution. Indeed, since $\rho_\delta(u)\le 2\delta|u|$, comparing the
objective at $\theta_{\varepsilon}=T^{(\delta,\varepsilon)}_{\P}(\theta')$ and at $c\in\P$ gives
\begin{equation}\label{eq:eps-bound}
\|\theta_{\varepsilon}-c\|_W^2\;\le\;\frac{2\delta}{\varepsilon}\sum_{i=1}^n w_i\,|\theta'_i-c_i| ,
\end{equation}
so that $T^{(\delta,\varepsilon)}_{\P}(\theta')\to c$ as $\delta\to0$ for fixed $\varepsilon$. The
regularization must therefore be small relative to $\delta$.

\noindent The natural choice $c=0$ has two drawbacks. First, the selected
solution depends on the origin of the time scale, which is arbitrary
(years BP, cal.\ BC/AD). Second, the map is not the identity on $\P$:
a sample that already satisfies the stratigraphic constraints is shrunk towards $0$, and the map is therefore not a projection.

\noindent We instead center the penalty at the weighted isotonic projection of
the sample, $c=T_\P(\theta')$. In the following proposition we prove
that  the resulting map is translation-equivariant, and leaves the points of
$\P$ unchanged.

\begin{proposition}\label{prop:recentred}
Let $\varepsilon>0$.  If the center is  $c=T_\P(\theta')$,  then 
$T^{(\delta,\varepsilon)}_\P$ satisfies  the following properties :  
\begin{enumerate}[(i)]
   \item $T^{(\delta,\varepsilon)}_\P(\theta')=\theta'$ for every $\theta'\in\P$;
  \item $T^{(\delta,\varepsilon)}_\P(\theta'+\tau\1)=T^{(\delta,\varepsilon)}_\P(\theta')+\tau\1$
        for all $\tau\in\R$.
\end{enumerate}
\end{proposition}

\begin{proof}
 Denote \[
  F_{\theta',c}(\theta)=\sum_{i=1}^nw_i\rho_{\delta}(\theta'_i-\theta_i)+\varepsilon\|\theta-c\|_W^2,
\]

(i) If $\theta'\in\P$, then $c = T_\P(\theta')= \arg\min_{\theta
                  \in \P}  \Vert \theta - \theta'\Vert_W^2 = \theta'$. Moreover
                 the objective $F_{\theta', \theta'}$  vanishes at $\theta'\in\P$,
so $\theta'$ is the minimizer by uniqueness.

(ii) Let $\theta' \in\mathbb{R}^n$ and $\tau\in\mathbb{R}$ be arbitrary. \\
First  we show that
$T_\P(\theta'+\tau\1)=T_\P(\theta')+\tau\1$. Denote
$J_{\theta'}(\theta)=\|\theta-\theta'\|_W^2$, we have for all  $\theta\in\R^n$,
\[
  J_{\theta'+\tau\1}(\theta+\tau\1)
  =J_{\theta'}(\theta).
\]
For all $\eta\in\P$, since $\P$ is invariant by
translation  $\eta-\tau\1\in \P$  and 
\[
  J_{\theta'+\tau\1}(\eta)=J_{\theta'}(\eta-\tau\1)\ \ge\
  J_{\theta'}(T_\P(\theta'))=J_{\theta'+\tau\1}(T_\P(\theta')+\tau \1).
\]
The minimizer of $J_{\theta'+\tau\1}$ over $\P$ is unique
thus  $T_\P(\theta'+\tau\1)=T_\P(\theta')+\tau\1$.
Furthermore 
$$F_{(\theta'+\tau\1,\,c+\tau\1)}(\theta+\tau\1)=F_{(\theta',c)}(\theta)$$
and for all $\eta\in\P$, 
\begin{align*}
 & F_{\theta'+\tau\1 , T_\P(\theta'+\tau\1) }(\eta)=  F_{\theta'+\tau\1
    , T_\P(\theta')+\tau\1 }(\eta)
  =  F_{\theta' , T_\P(\theta' ) }(\eta- \tau\1) \\
  & \geq  F_{\theta' ,
    T_\P(\theta' ) }( T^{(\delta,\varepsilon)}_\P(\theta'))  = F_{\theta' +\tau\1,
    T_\P(\theta' ) +\tau\1}( T^{(\delta,\varepsilon)}_\P(\theta')
  +\tau\1) = F_{\theta' +\tau\1,
    T_\P(\theta' +\tau\1)}( T^{(\delta,\varepsilon)}_\P(\theta')
  +\tau\1)
\end{align*}
By uniqueness of the minimizer of  $ F_{\theta' +\tau\1,
  T_\P(\theta' +\tau\1)}$ over $\P$, we get $T^{(\delta,\varepsilon)}_\P(\theta'+\tau\1)=T^{(\delta,\varepsilon)}_\P(\theta')+\tau\1$. 
\end{proof}
\noindent The Huber loss depends on the hyperparameter $\delta$. As
$\delta \to \infty$, the resulting objective function $H_\delta$
approaches the Isotonic Regression objective function $Q$ defined in \eqref{eq:IR}. The following proposition shows that when  $\delta$ exceeds an explicit and data-driven threshold, the two objective functions share a common minimizer. 

\begin{proposition} \label{prop:delta_threshold}
  Let $r = \theta' - T_\P(\theta')$, be the residual of the Isotonic Regression solution \eqref{eq:IR}. 
  
  \begin{enumerate}[(i)] 
\item If $\norm{r}_\infty \leq \delta$, then $T_\P(\theta')\in T^{(\delta)}_\P(\theta')$.
\item If $\norm{r}_\infty \leq \delta$ and $c=T_\P(\theta')$, then
$T^{(\delta,\varepsilon)}_\P(\theta')=T_\P(\theta')$ for every $\varepsilon>0$.
\item If $\norm{r}_\infty < \delta$, then $T^{(\delta)}_\P(\theta')=\{T_\P(\theta')\}$.
\end{enumerate}
\end{proposition}

\begin{proof} ~ \label{proof:threshold_huber}  $H_\delta$ is a $\mathcal{C}^1$ function and its gradient reads \begin{equation}
  \partial_{\theta_i} H_\delta (\theta) = -2w_i
\min \bigg(1, \frac{\delta}{\mid \theta'_i - \theta_i \mid}\bigg) (\theta'_i - \theta_i) \quad \forall \; i.
\end{equation}  \\ 
\noindent (i):  Since $|r_i|\le\delta$, we have  $\partial_{\theta_i}H_{\delta}(T_\P(\theta'))= -2w_i
r_i $ and 
\begin{equation} \label{eq:a1}
  \nabla H_{\delta}(T_\P(\theta'))=-2Wr=\nabla Q(T_\P(\theta')), \end{equation}  
where W is the matrix defined in \eqref{eq:W}. 
    
\noindent As $T_\P(\theta')$ minimizes $Q$ over $\P$,
the first-order condition (see \cite{boyd_convex_2023}) gives $$\langle\nabla
Q(T_\P(\theta')),\theta-T_\P(\theta')\rangle\ge0$$ for all $\theta\in\P$ and 
using \eqref{eq:a1} the same inequality holds for $H_{\delta}$

$$\langle\nabla
H_\delta(T_\P(\theta')),\theta-T_\P(\theta')\rangle\ge0$$ for all $\theta\in\P$
and $T_\P(\theta')$ minimizes $H_{\delta}$ over $\P$ i.e. $T_\P(\theta')\in T^{(\delta)}_\P(\theta')$.

\noindent (ii):   According to (i),   $H_{\delta}(\theta)\ge H_{\delta}(T_\P(\theta'))$ for all $\theta\in\P$. Hence, for
$\theta\in\P\setminus\{T_\P(\theta')\}$,
\[
H_{\delta}(\theta)+\varepsilon\|\theta-T_\P(\theta')\|_W^2>H_{\delta}(T_\P(\theta'))
=H_{\delta}(T_\P(\theta'))+\varepsilon\|T_\P(\theta')-T_\P(\theta')\|_W^2 ,
\]
so $T_\P(\theta')$ is the unique minimizer of the regularized objective.

\noindent (iii):   Let $\hat\theta\in
T^{(\delta)}_\P(\theta')$, $d=\hat\theta-T_\P(\theta')$ and 
$t\in[0,1]$ thus, as $\P$ is a convex set, we have 
$$\theta_t=T_\P(\theta')+t\,d= (1-t) T_\P(\theta') + t \hat \theta \in\P.$$ 
By (i)  and $H_\delta$ convexity, $$H^* := \min_\P H_{\delta} \le H_{\delta}(\theta_t)\le(1-t)H_{\delta}(T_\P(\theta'))+tH_{\delta}(\hat\theta)=H^*, $$ thus
$H_{\delta}(\theta_t)=H^*$ for all $t\in[0,1]$. \\ 
Assume $d\ne0$ and set
\[
t_0=\min\Bigl(1,\ \min_{1\le i\le n}\frac{\delta-|r_i|}{\|d\|_\infty}\Bigr)>0 .
\]
For $t\in[0,t_0]$ and every $i$, $|\theta'_i-(\theta_t)_i|=|r_i-t\,d_i|\le|r_i|+t\|d\|_\infty\le\delta$, so all
residuals are in the quadratic regime and
\[
H_{\delta}(\theta_t)=Q(\theta_t)=Q(T_\P(\theta'))-2t\sum_{i=1}^n w_ir_id_i+t^2\sum_{i=1}^n w_id_i^2 .
\]
This polynomial is constant on $[0,t_0]$, hence $\sum_iw_id_i^2=0$, i.e.\ $d=0$. This is a contradiction and $T_\P^{(\delta)}(\theta') = \{T_\P(\theta')\}$. 

\end{proof}

\subsection{Monotone Baseline and Sparse Spikes}
We introduce an alternative objective function in which the signal decomposes as the sum of a monotone baseline (MBL) $\theta \in \mathcal{P}$ and a sparse perturbation
$s \in \mathbb{R}^n$, called spikes. The latter accounts for deviations when faced with outliers. The associated optimization reads \begin{align} \label{eq:MBL}
	T_\P^{(\eta)} (\theta')&= \underset{s \in \R^n \ \theta \in \P}{\arg\min} \sum_{i=1}^{n} w_i (\theta'_i - \theta_i-s_i)^2 + \eta \norm{s}_1 \quad \forall \ \theta' \ \in \R^n  \nonumber \\ 
  & := \underset{s \in \R^n \ \theta \in \P}{\arg\min} H_\eta(\theta, s) \quad \forall \ \theta' \ \in \R^n. 
\end{align}
This explicit decomposition aims to disentangle the underlying monotone component $\theta$ from the observed measurements $\theta'$. The resulting deviations $s$ are expected to exhibit large spikes at contaminated observations, relative to the deviations associated with non-contaminated measurements.
\begin{proposition}\label{prop:mbl_huber}
  The MBL optimization problem \eqref{eq:MBL}
  is equivalent to a weighted Huber isotonic regression \eqref{eq:huber_map} with
  sample-wise threshold $\eta_i := \eta / (2w_i)$. Specifically,
  the inner minimization admits the closed-form solution
  \begin{align}\label{eq:sstar}
    s_i^* &=
    \begin{cases}
      r_i - \operatorname{sign}(r_i)\,\dfrac{\eta}{2w_i}
        & \text{if } |r_i| > \dfrac{\eta}{2w_i}, \\[6pt]
      0 & \text{if } |r_i| \leq \dfrac{\eta}{2w_i},
    \end{cases} \\
    &= \operatorname{sign}(r_i)\big(\mid r_i \mid - \dfrac{\eta}{2w_i}\big)_+
  \end{align}
  where $r_i := \theta'_i - \theta_i$, and substituting the spikes solution~\eqref{eq:sstar}
  back into~\eqref{eq:MBL} yields
  \begin{equation}\label{eq:mbl_huber}
    \min_{\theta \in \mathcal{P}}\;
    \sum_{i=1}^{n} w_i\,\rho_{\eta_i}(r_i),
  \end{equation}
  where $\rho_{\eta_i}$ denotes the Huber loss with threshold
  $\eta_i = \eta/(2w_i)$.
\end{proposition}

\begin{proof}
  Fix $\theta$ and consider the scalar problem in $s_i$. Define
  \begin{equation*}
    G(s_i) := w_i(r_i - s_i)^2 + \eta|s_i|
    =
    \begin{cases}
      w_i(r_i - s_i)^2 + \eta s_i & \text{if } s_i \geq 0, \\
      w_i(r_i - s_i)^2 - \eta s_i & \text{if } s_i \leq 0.
    \end{cases}
  \end{equation*}
  Since $G$ is convex and coercive, it attains its minimum. For
  $s_i \neq 0$, $G$ is differentiable with
  \begin{equation*}
    G'(s_i) =
    \begin{cases}
      -2w_i(r_i - s_i) + \eta & \text{if } s_i > 0, \\
      -2w_i(r_i - s_i) - \eta & \text{if } s_i < 0.
    \end{cases}
  \end{equation*}
  Setting $G'(s_i) = 0$ on each branch gives the candidate
  $s_i = r_i \mp \eta/(2w_i)$. To verify these are global minimizers,
  we check the sign of $G'$ on each branch. For $s_i > 0$,
  \begin{equation*}
    G'(s_i) \leq 0 \iff s_i \leq r_i - \frac{\eta}{2w_i},
  \end{equation*}
  so $G$ is decreasing then increasing on $(0,+\infty)$, confirming
  $s_i^* = r_i - \eta/(2w_i) > 0$ whenever $r_i > \eta/(2w_i)$.\\ 
  The case $s_i < 0$ is symmetric and yields $s_i^* = r_i + \eta/(2w_i) < 0$ whenever $r_i < -\eta/(2w_i)$.\\
  For $|r_i| \leq \eta/(2w_i)$, the subdifferential condition at
  $s_i = 0$ requires $0 \in \partial G(0) = \{-2w_i r_i\} +
  \eta[-1,1]$, i.e.\ $|2w_i r_i| \leq \eta$, which holds exactly
  in this regime. Hence $s_i^* = 0$.

 \noindent  Substituting the spikes solutions into $G(s_i^*)$ gives \begin{equation*}
    G(s_i^*) =
    \begin{cases}
      w_i r_i^2 & \text{if } |r_i| \leq \eta_i, \\
      \eta|r_i| - \tfrac{\eta^2}{4w_i}
        = w_i\!\left(2\eta_i|r_i| - \eta_i^2\right)
        & \text{if } |r_i| > \eta_i,
    \end{cases}
  \end{equation*} with $\eta_i = \eta/(2w_i)$, which is precisely $r_i \mapsto w_i\,\rho_{\eta_i}(r_i)$. Summing over $i$ and minimizing over $\theta \in \mathcal{P}$ yields~\eqref{eq:mbl_huber}.
\end{proof}
\noindent Since each observation has its own effective threshold, \(\eta_i = \eta/(2w_i)\), larger weights result in a smaller inlier region \(\lbrace r \in \R ;\ \mid r \mid < \eta_i \rbrace \). Thus, more heavily weighted observations enter the robust (linear) regime of the Huber loss for smaller residuals. \\
The non-uniqueness issue encountered in the standard Huber isotonic regression problem also arises in this setting. We therefore employ the Tikhonov-regularized formulation for this projection as well.
  \begin{equation}\label{eq:mbl_tikh}
    T_\P^{(\eta, \varepsilon)} (\theta')= \arg\min_{\theta \in \mathcal{P}}\;
    \sum_{i=1}^{n} w_i \rho_{\eta_i}(\theta'_i - \theta_i)
    + \varepsilon \norm{\theta-c}_W^2,
  \end{equation}
  with the corresponding spike estimate
  \begin{equation}\label{eq:sstar_tikh}
    s_i = \operatorname{sign}(r_i) \!\left( \mid r_i \mid - \eta_i\right)_+,
  \end{equation}
where $r := \theta' - T_\P^{(\eta, \varepsilon)} (\theta')$, is the vector of residuals. The equations \eqref{eq:mbl_tikh} and \eqref{eq:sstar_tikh}  define the unique solution of
\begin{equation}\label{eq:mbl-joint}
  H_{\eta,\varepsilon}(\theta,s)=\sum_{i=1}^nw_i(\theta'_i-\theta_i-s_i)^2+\eta\|s\|_1
  +\varepsilon\|\theta-c\|_W^2 ,
\end{equation}

\begin{remark}
  The MBL and Huber maps belong to the same family of robust maps defined by
\begin{equation} \label{eq:robust_family}
  \min_{\theta\in\P,\ s\in\R^n}\ \sum_{i=1}^n w_i(\theta'_i-\theta_i-s_i)^2+\sum_{i=1}^n\lambda_i|s_i|,
\end{equation}
where $\lambda=(\lambda_1,\dots,\lambda_n)\in(0,\infty)^n$. 
By the argument of Proposition \ref{prop:mbl_huber}, minimizing in $s$ yields weighted Huber isotonic
regression with hyperparameters $\tau_i=\lambda_i/(2w_i)$, and spikes 
$s_i^\star=\operatorname{sign}(r_i)\,(|r_i|-\tau_i)_+$.  Our two choices correspond to 
\begin{itemize}
\item $\lambda_i=2\delta w_i$, i.e.\ $\tau_i\equiv\delta$: the Huber map;
\item $\lambda_i\equiv\eta$, i.e.\ $\tau_i=\eta/(2w_i)$: the MBL map.
\end{itemize}

\noindent Note that this family \eqref{eq:robust_family} also contains the map defined by the Huber loss evaluated over standardized residuals i.e. the objective function is $\sum_{i=1}^{n} \rho_\delta(\sqrt{w_i} (\theta_i' - \theta_i))$. This means $\lambda_i = 2 \delta \sqrt{w_i}$ and  $\tau_i = \delta (w_i)^{-1/2}$. 
\end{remark}
\noindent The uniqueness analysis in Proposition \ref{prop:huber_unic} and
Corollary \ref{cor:unique-root} extends to the whole family~\eqref{eq:robust_family}. On a
block $B_o$, the stationarity function becomes
\[
h_\tau(\theta)=-2\sum_{i\in B_o}w_i\min\Bigl(1,\frac{\tau_i}{|\theta'_i-\theta|}\Bigr)(\theta'_i-\theta),
\qquad \tau_i=\frac{\lambda_i}{2w_i}.
\]

\begin{proposition}\label{prop:gap-balance}
The solution set of $h_\tau(\theta)=0$ contains a non-trivial interval if and only if
there is a partition $B_o=A\sqcup C$, with $A$ and $C$ non-empty, such that
\[
\max_{i\in A}(\theta'_i+\tau_i)<\min_{j\in C}(\theta'_j-\tau_j)
\qquad\text{and}\qquad
\sum_{i \in A} \lambda_i= \sum_{i \in C} \lambda_i .
\]
\end{proposition}

\begin{proof}
The proof of Proposition \ref{prop:huber_unic} applies with $\delta$ replaced by $\tau_i$: writing
$h_\tau=\sum_{i\in B_o}\varphi_i$ with
$\varphi_i(\theta)=2w_i\max\bigl(-\tau_i,\min(\tau_i,\theta-\theta'_i)\bigr)$, each $\varphi_i$ is
non-decreasing, strictly increasing on $[\theta'_i-\tau_i,\theta'_i+\tau_i]$, and equal to
$\pm\lambda_i$ outside this interval. The only change is that the saturated
contributions are $\pm\lambda_i$ instead of $\pm2\delta w_i$, so that the balance
condition is $\sum_{i \in A} \lambda_i =\sum_{i \in C} \lambda_i$, and the gap condition involves the
sample-wise thresholds $\tau_i$.
\end{proof}

\noindent For the Huber map ($\lambda_i=2\delta w_i$), the balance condition is
$\sum_{i \in A} w_i=\sum_{i \in A} w_i$, and the gap condition forces $A$ to consist of the smallest values,
which recovers Proposition~\ref{prop:huber_unic}. For the MBL map ($\lambda_i\equiv\eta$), it reduces
to $|A|=|C|$ and does not depend on the weights. \\
For the Huber loss on standardized residuals ($\lambda_i=2\delta\sqrt{w_i}$), it reads
$\sum_{i\in A}\sqrt{w_i}=\sum_{j\in C}\sqrt{w_j}$.

\begin{remark}
When all the weights are equal, the Huber and MBL maps coincide with $\eta=2w\delta$,
both with and without the Tikhonov term~\eqref{eq:huber_reg_map}.). In our applications, however, the
weights are estimated from the MCMC output of the Unconstrained model
(Section~\ref{sec:hyperparam}), and the balance condition $\sum_{i \in A} w_i=\sum_{i \in C} w_i$ of the Huber map is not
satisfied in practice, so that Tikhonov regularization is not needed to ensure
uniqueness of the Huber map. For the MBL map, the balance condition $|A=|C|$ of
Proposition~\ref{prop:gap-balance} does not depend on the weights: non-uniqueness
cannot be excluded, and Tikhonov regularization is required.
\end{remark}

\noindent Similar to the Huber case, the hyperparameter $\eta$ has an explicit threshold defined in the following proposition. 
\begin{proposition} \label{prop:eta_thereshold}
If $\eta\ge\|2W (\theta' - T_\P(\theta'))\|_\infty$ and  $c=T_\P(\theta')$, then $(T_\P(\theta'), 0)$ is the unique solution of the optimization problem \eqref{eq:mbl-joint}.
\end{proposition}
\begin{proof}
  Immediate adaptation of the Proof of  Proposition \ref{prop:delta_threshold}. 
\end{proof}

\subsection{Log-hyperbolic cosine}
Smooth approximations of the Huber loss arise naturally within the family of generalized Huber losses introduced by \cite{gokcesu_generalized_2021}. In particular, this framework provides a class of smooth robust loss functions that combine quadratic behavior near the origin with linear growth in the tails, mirroring the key properties of the Huber loss \eqref{eq:huber_loss}. These function have the following form \begin{equation*} 
  L(r) = f^{-1}\bigl(f(r) + f(-r)\bigr), \qquad r \in \mathbb{R},
\end{equation*}
where \(f\) is smooth and strictly increasing, with a finite limit as \(x\to-\infty\), and satisfies \(\lim_{x\to+\infty}f(x)=+\infty\).
\noindent A canonical choice is $f(r) = \exp(ar) + b$, which yields
\begin{equation*} 
  L(r) = \frac{1}{a}\log\bigl(\exp(ar) + \exp(-ar) + b\bigr) \quad a >0.
\end{equation*}
Setting $a = 1/\kappa$ with $\kappa > 0$ and $b = 0$, we can recover the \emph{log-cosh loss}. In this case, the resulting loss is obtained from $L$ by a translation by $- \kappa \log(2)$ and rescaling by $2 \kappa$ of $L$. 
\begin{equation}\label{eq:logcosh}
  \rho_\kappa(r) = 2\kappa^2\log\cosh\!\left(\frac{r}{\kappa}\right),
  \qquad r \in \mathbb{R},
\end{equation}
Its first two derivatives are \begin{align*}
  \rho'_\kappa(r)   &= 2 \kappa \tanh\!\left(\frac{r}{\kappa}\right),
   \\
  \rho''_\kappa(r)  &= 2\operatorname{sech}^2\!\left(\frac{r}{\kappa}\right).
\end{align*}
The corresponding asymptotic behaviors  \begin{equation*}
	\rho_\kappa(r) = \begin{cases}
          r^2 + O(r^4 /\kappa^2) & \mid r \mid \ll  \kappa \\
		
		2 \kappa \mid r \mid - 2\kappa^2 \log(2) + O(\exp(-2\mid r\mid / \kappa)) & \mid r \mid \gg \kappa,
	\end{cases}
\end{equation*}
show the quadratic behavior near the origin and the linear growth in the tails.
\begin{proposition} \label{prop:logcosh_map}
  Let $\kappa > 0$. The map \begin{equation} \label{eq:logcosh_map}
	T_\P^{(\kappa)}(\theta') = \arg\min_{\theta \in \P} \sum_{i=1}^{n} w_i \rho_\kappa(\theta'_i - \theta_i) =: \arg\min_{\theta \in \P} H_\kappa(\theta) \quad \forall \ \theta' \in \R^n,
\end{equation}
is well-defined and admits a unique solution. 
\end{proposition}
\begin{proof}
  The loss function $\rho_\kappa$, admits a second derivative $\rho''_\kappa > 0$, thus it is strictly convex and a sum of strictly convex functions is still a strict convex function. Moreover, the objective function is coercive since $\rho_\kappa(r) \geq 2\kappa \mid r \mid - 2\kappa^2\log2\underset{|r|\rightarrow + \infty}{ \longrightarrow} \infty$. Together, they guarantee the existence of a unique minimizer. 
\end{proof}

\begin{proposition} \label{prop:block_sol_logcosh}
  Let $B_1 \cup \cdots \cup B_K$ be the partition of the optimal solution $T_\P^{(\kappa)}(\theta')$. Over each block $B_o, \; o \in \{1, \cdots, K\}$, the KKT stationarity conditions reduce to the scalar equation \begin{equation} \label{eq:block_sol_logcosh}
    h_\kappa(\theta_o) = 0,
  \end{equation} 
  where \[h_\kappa(t) = -2 \kappa \sum_{i \in B_o} w_i \tanh(\frac{(\theta'_i-t)}{\kappa}) \quad \forall t \in \R.\]
  The solution $\theta_o$ exists, is unique. Moreover, 
\[\theta_o \in \lbrack \min_{i \in B_o} \theta'_i, \max_{i \in B_o} \theta'_i \rbrack\]
\end{proposition}
\begin{proof}
  The function $h_\kappa$ has a positive derivative \[h'_\kappa(t) =  2 \sum_{i \in B_o} w_i sech^2(\frac{\theta'_i- t}{\kappa}) > 0 \ \forall t \in \R, \] hence is a strictly increasing,  continuous function and \begin{equation*}
    \lim_{t \rightarrow - \infty} h_\kappa(t) = -2\kappa \sum_{i \in B_o} w_i \quad 
		\lim_{t \rightarrow + \infty} h_\kappa(t) = 2\kappa \sum_{i \in B_o} w_i,
  \end{equation*}
  The existence and uniqueness of the solution \eqref{eq:block_sol_logcosh} is immediate with the intermediate value theorem. 
  Finally, \[h_\kappa(\min_{i \in B_o} \theta'_i) \leq 0 \quad \text{and} \quad h_\kappa(\max_{i \in B_o} \theta'_i) \geq 0.  \]
\end{proof}
\noindent Unlike the Huber map, no finite $\kappa$ yields $T_\P^{(\kappa)} = T_\P$ in general, since $\rho_\kappa$ is nowhere exactly quadratic, Proposition~\ref{prop:blocklogch} instead quantifies the approximation.
\begin{lemma}\label{lem:implicit}
  Let $r = \theta' - T_\P^{(\kappa)} (\theta')$, be the residual of the log-hyperbolic cosine solution, then \begin{equation*}
    T_\P^{(\kappa)} (\theta') = T_\P(\theta' - \xi ),
  \end{equation*}
  where $\xi $ is the vector for which each coordinate is $r_i - \kappa \tanh(r_i/ \kappa)$.
\end{lemma}
\begin{proof}
    Let $z_i=\theta'_i- (r_i - \kappa \tanh(r_i/ \kappa))= T_\P^{(\kappa)} (\theta')_i +\kappa\tanh(r_i/\kappa)$ and $G(\theta)=\sum_i w_i(z_i-\theta_i)^2$. \\
Then $\nabla G(T_\P^{(\kappa)} (\theta'))=\nabla H_\kappa(T_\P^{(\kappa)} (\theta'))$,
so the first-order condition of $H_\kappa$ at $T_\P^{(\kappa)} (\theta')$ is that of the convex function $G$,
hence $T_\P^{(\kappa)} (\theta')$ minimizes $G$ over $\P$, i.e.\ $T_\P^{(\kappa)} (\theta')=T_\P(z)$.
\end{proof}
\begin{remark}\label{rem:spike_existence}
The argument of Lemma \ref{lem:implicit} applies to the Huber loss, every
$\hat\theta\in T^{(\delta)}_{\P}(\theta')$ satisfies
\[
\hat\theta=T_\P\big(\theta'-s\big),\qquad s_i= \operatorname{sign}(r_i)\,(|r_i|-\delta)_+,\quad r=\theta'-\hat\theta.
\]
Therefore the  robust solution is the isotonic projection cleaned from its spikes.
For the log-hyperbolic cosine, this lemma  gives the smooth analogue
$s_i=r_i-\kappa\tanh(r_i/\kappa)$. 
\end{remark}

\begin{proposition} \label{prop:blocklogch}
Let $\kappa>0$ and $D(\theta')=\max_i\theta'_i-\min_i\theta'_i$. Then
\[
\big\|T^{(\kappa)}_{\P}(\theta')-T_\P(\theta')\big\|_W\;\le\;
\frac{D(\theta')^3}{3\kappa^2}\Big(\sum_{i=1}^n w_i\Big)^{1/2}.
\]
In particular $T^{(\kappa)}_{\P}(\theta')\to T_\P(\theta')$ as $\kappa\to\infty$, at rate
$O(\kappa^{-2})$. 
\end{proposition}
\begin{proof}
By Lemma \ref{lem:implicit} , $T^{(\kappa)}_{\P}(\theta')=T_\P(\theta'-\xi)$ with
$\xi_i=r_i-\kappa\tanh(r_i/\kappa)$ and $r=\theta'-T^{(\kappa)}_{\P}(\theta')$. \\
Since $0\le x-\tanh x\le x^3/3$ for $x\ge0$ and $x \mapsto x-\tanh(x) $ is odd, we have \[|\xi_i|\le|r_i|^3/(3\kappa^2).\] 
By Proposition \ref{prop:block_sol_logcosh}, every block value of $T^{(\kappa)}_{\P}(\theta')$ lies in
$[\min_j\theta'_j,\max_j\theta'_j]$, hence $|r_i|\le D(\theta')$. \\
Finally, $T_\P$ is the $\|\cdot\|_W$-projection onto the closed convex set $\P$, hence $1$-Lipschitz for $\|\cdot\|_W$, and \[\|T_\P(\theta'-\xi)-T_\P(\theta')\|_W\le\|\xi\|_W\le\max_i|\xi_i|\,(\sum_iw_i)^{1/2}.\]
\end{proof}

\noindent The Huber loss becomes linear for $|r|>\delta^\star$ ($\delta^\star$ being the collapse threshold fixed in \ref{prop:delta_threshold}), where its derivative reaches $2\delta^\star$; the
log-cosh derivative $2\kappa\tanh(r/\kappa)$ only approaches $2\kappa$. Using the saturation
tolerance $\nu$, we require a residual $|r|=\delta^\star$ to be in the saturation regime,
$|\tanh(\delta^\star/\kappa)|=1-\nu$:
\begin{equation}\label{eq:kappa_star}
  \kappa^\star=\frac{\delta^\star}{\operatorname{arctanh}(1-\nu)},\qquad
  \operatorname{arctanh}(1-\nu)=\frac12\log\frac{2-\nu}{\nu}.
\end{equation}

\noindent We next describe the algorithms used to compute each of the projection maps introduced above.

\section{Computational approach for map solving} \label{sec:computational}

\noindent In the remainder of the paper, the Huber and MBL maps are
understood to refer to their Tikhonov-regularized formulations. For
simplicity of notation, we omit the regularization parameter
\(\varepsilon\) and denote these maps by \(T_{\mathcal P}^{(\delta)}\)
and \(T_{\mathcal P}^{(\eta)}\), respectively. 

\noindent The projection maps considered in this work lead to convex optimization problems over the isotonic cone. Although generic constrained optimization methods could be used to solve these problems, the specific structure of the isotonic constraints allows for more specialized algorithms.

\noindent The Generalized Isotonic Recursive Partitioning (GIRP) algorithm of \citet{luss2014generalized} extends the Isotonic Recursive Partitioning (IRP) algorithm of \citet{luss2012efficient}. While IRP addresses least-squares isotonic regression under a partial order through recursive partitioning of the constraint graph, GIRP replaces the group mean with a group-wise minimizer of the objective function. This generalization allows GIRP to handle separable, convex, differentiable loss functions under arbitrary partial orders, provided that the corresponding subproblems can be solved appropriately. Its optimality guarantee relies on the exact solution of these subproblems.

\noindent More recently, \citet{wang_efficient_2022} introduced Sequential Block Merging (SBM), a primal-dual block-merging algorithm that decomposes KKT conditions into local systems. SBM accommodates arbitrary DAG constraints and separable, convex, differentiable objectives satisfying the assumptions stated in their paper, including the solvability of the stationarity equation for each individual term. Under these assumptions, their Theorem 2 establishes convergence to an optimal solution.

\noindent In this work, we adopt SBM to solve the robust projection problems arising from our projection-based approach. This choice allows us to investigate its application beyond the quadratic regression setting considered in the original SBM paper, particularly for the Huber loss and related robust losses. We introduce a regularization term that ensures uniqueness of the solution, thereby providing a well-defined reference for evaluating the algorithms. This choice is particularly relevant when comparing SBM with GIRP, whose behavior may be affected by non-unique minimizers (see \cite{won2024correctness}).

\subsection{Adaptation of the SBM for the Huber loss}
\begin{proposition} \label{prop:certif_huber_sbm}
  The regularized objective function $H_{(\varepsilon, \delta)}$ is convex, differentiable, separable, i.e, it decomposes into a sum of functions \[H_{(\varepsilon, \delta)} (\theta) = \sum_{i=1}^{n} f_{\theta'_i, c_i}(\theta_i),\]
where $f_{\theta'_i, c_i}(x) = w_i \rho_\delta(\theta'_i- x) + \varepsilon w_i (x-c_i)^2$, and the solution of $f'_{\theta'_i, c_i} (x^*) = 0$ exists and reads \[x^* = \begin{cases}
	\frac{\theta'_i + \varepsilon c_i}{(1+\varepsilon)} & \text{if } \mid \theta'_i-c_i \mid \leq \frac{1+\varepsilon}{\varepsilon} \delta \\
	c_i + \frac{\delta}{\varepsilon} & \text{if } \theta_i'-c_i > \frac{1+\varepsilon}{\varepsilon}\delta \\
	c_i - \frac{\delta}{\varepsilon} & \text{else}.
\end{cases}\]
\end{proposition}

\begin{proof}
	The derivative reads \begin{equation*}
		f'_{\theta_i', c_i}(x) = -2w_i \min\bigg(1, \frac{\delta}{\mid \theta_i'- x \mid}\bigg) (\theta_i'- x)+ 2 \varepsilon w_i (x-c_i),
	\end{equation*}
	It is strictly increasing function and \[\begin{cases}
		\lim_{x \to -\infty} f'_{\theta'_i, c_i}(x) = -\infty \\
		\lim_{x \to +\infty} f'_{\theta'_i, c_i}(x) = +\infty.
	\end{cases}\]
The existence and uniqueness of the solution is immediate with the intermediate value theorem
\end{proof}
\noindent In the remainder of the paper, we set the regularization center to the projection \(c=T_{\mathcal P}(\theta')\), for all \(\theta'\in\mathbb{R}^n\).
\begin{proposition} The SBM algorithm applied to the objective function $H_{(\varepsilon, \delta)}$ converges to its unique minimizer.
\end{proposition}
\begin{proof}
	The objective function verifies the Assumption 1 presented in \cite{wang_efficient_2022}, hence the method is convergent. 
\end{proof}

\noindent The unique solution to the $T_\P^{(\delta)}$ projection can be retrieved as stated before with the SBM, however one needs to solve the decomposed KKT equation at an arbitrary block $B_o$. 
We set, for a fixed $\theta \in \mathbb{R}$, the following block partition
\begin{equation*}
\begin{cases}
U(\theta)=\{i\in B_o \mid \theta<\theta'_i-\delta\} & \text{(upper plateau)}\\
M(\theta)=\{i\in B_o \mid \theta'_i-\delta\le\theta<\theta'_i+\delta\} & \text{(linear ramp)}\\
D(\theta)=\{i\in B_o \mid \theta'_i+\delta\le\theta\} & \text{(lower plateau)}
\end{cases}
\end{equation*}
so that $B_o = U(\theta) \sqcup M(\theta) \sqcup D(\theta)$ for every $\theta$.
The block stationarity equation, under Thikhonov regularization reads
\begin{align} \label{eq:huber_block_eq}
	h_\delta(\theta) &= -2 \sum_{i \in B_o} w_i \min\!\left(1, \frac{\delta}{|\theta'_i - \theta|} \right)(\theta'_i - \theta) + 2\varepsilon \sum_{i \in B_o} w_i \, (\theta - T_\P(\theta')_i) = 0, \qquad \forall \theta \in \mathbb{R}.
\end{align}
Since the weighted terms  \begin{equation} \label{eq:huber_weights}
  w_{0, i}^{(\delta)}:= \min\left(1, \dfrac{\delta}{|\theta'_i - \theta|}\right) \quad \forall i,
\end{equation} decide whether the $i-$th component is in the linear or quadratic regime, each summand is a plateau for $i \in U(\theta) \cup D(\theta)$ and a linear ramp for $i \in M(\theta)$. Splitting the sum over the three sets gives, for fixed block sets,
\begin{align*}
	h_\delta(\theta) &= -2\delta \bigg[ \sum_{j \in U(\theta)} w_j - \sum_{j \in D(\theta)} w_j \bigg] - 2 \sum_{j \in M(\theta)} w_j (\theta_j' - \theta) + 2\varepsilon \Big(\sum_{j \in B_o} w_j\Big) \theta - 2\varepsilon \Big(\sum_{j \in B_o} w_j T_\P(\theta')_j \Big) \nonumber\\
	&= \underbrace{-2\delta \bigg[ \sum_{j \in U(\theta)} w_j - \sum_{j \in D(\theta)} w_j \bigg] - 2 \sum_{j \in M(\theta)} w_j \theta_j' - 2\varepsilon \Big(\sum_{j \in B_o} w_j T_\P(\theta')_j \Big)}_{=: \; c(\theta)} \; + \; \underbrace{2 \bigg[ \sum_{j \in M(\theta)} w_j + \varepsilon \sum_{j\in B_o} w_j \bigg]}_{=: \; m(\theta)} \, \theta.
\end{align*}
On any interval where $U(\theta)$, $M(\theta)$, $D(\theta)$ stay constant, $h_\delta$ is exactly affine: $h_\delta(\theta) = c(\theta) + m(\theta)\,\theta$ with $c(\theta)$, $m(\theta)$ constant on that interval. Thanks to the Tikhonov term, the slope $m(\theta) \geq 2\varepsilon \sum_{j \in B_o} w_j> 0$, so $h_\delta$ is strictly increasing and admits a unique root $\theta^\star$.

\noindent The sets $U(\theta), M(\theta), D(\theta)$ only change at the $2\mid B_o \mid$ kink points
\[
\begin{cases}
	l_i = \theta_i' - \delta, & i \in B_o \quad \text{(entry: $i$ moves $U \to M$)} \\
	L_i = \theta_i' + \delta, & i \in B_o \quad \text{(exit: $i$ moves $M \to D$)}.
\end{cases}
\]
As $\theta$ increases past $l_i$, sample $i$ switches from $U(\theta)$ to $M(\theta)$; as $\theta$ increases past $L_i$, sample $i$ switches from $M(\theta)$ to $D(\theta)$. Let us denote these ordered kink points as follows \[t_1 \leq \cdots \leq t_{2\mid B_o \mid},\]
then for arbitrary starting point $t_0 < \min_{i \in B_o}(l_i) := t_1$,  $U(t_0) = B_o$ and $M(t_0) = D(t_0) = \emptyset$. Hence,
	\[
		c(t_0) = -2\delta \sum_{j \in B_o} w_j  -2\varepsilon\sum_{j \in B_o} w_j  T_\P(\theta')_j , \qquad m(t_0) = 2\varepsilon \sum_{j \in B_o}w_j, 
	\]
Alternatively, for an arbitrary point $t_{2\mid B_o \mid +1} > \max_{i \in B_o}(L_i) = t_{2\mid B_o \mid}$, $U(t_{2\mid B_o \mid+1}) = M(t_{2\mid B_o \mid+1}) = \emptyset$ and $D(t_{2\mid B_o \mid+1}) = B_o$, hence, 
	\[
		c(t_{2\mid B_o \mid+1}) = 2\delta \sum_{j \in B_o} w_j  -2\varepsilon\sum_{j \in B_o} w_j  T_\P(\theta')_j , \qquad m(t_{2\mid B_o \mid +1}) = 2\varepsilon \sum_{j \in B_o}w_j. 
	\]

 For $k \in \{0, \cdots, 2\mid B_o \mid +1\}$, the function $h_\delta$ is linear over each interval $\lbrack t_k, t_{k+1} )$ and its expression reads
\[h_\delta(\theta) = h_\delta(t_k) + m(t_k)(\theta-t_k) \quad \forall \theta \in \lbrack t_k, t_{k+1}).\]
Thanks to $h_\delta$ strict monotonicity, the solution exists and is unique. Therefore there is a unique $k^*$ such that $h_\delta(t_{k^*}) \leq 0 < h_\delta(t_{k^*+1})$ and the solution of \eqref{eq:huber_block_eq} is   \[\theta^{*} = t_{k^*} - \dfrac{h_\delta(t_{k^*})}{m(t_{k^*})} < t_{k^*+1}. \]
Algorithm~\ref{alg:piecewise_root} describes the sweeping process with complexity $O(|B_o|\log(|B_o|))$ due to the sorting step. 

\begin{algorithm}
\caption{Root finding by a piecewise-affine sweep}
\label{alg:piecewise_root}
\begin{algorithmic}[1]
\REQUIRE Block $B_o$, function $h_\delta$

\STATE Sort the $2|B_o|$ breaking points:
       $t_1 < \cdots < t_{2|B_o|}$
\STATE Choose $t_0 < \min_{i\in B_o} l_i$ and $t_{2|B_o|+1} > \max_{i \in B_o} L_i$
\STATE Evaluate $h_\delta(t_0)$ and initialize the slope $m(t_0)$

\FOR{$k=1,\ldots,2|B_o|+1$}
    \STATE Compute the candidate root:
    \[
    \theta^{\mathrm{cand}}
    \gets t_{k-1}
    - \frac{h_\delta(t_{k-1})}{m(t_{k-1})}
    \]
    \IF{$\theta^{\mathrm{cand}} \leq t_k$}
        \STATE $\theta^\star \gets \theta^{\mathrm{cand}}$
         \RETURN $\theta^\star$
    \ELSE
        \STATE Advance to $t_k$ and evaluate:
        \[
        h_\delta(t_k)
        \gets h_\delta(t_{k-1})
        +m(t_{k-1})(t -t_{k-1}),
        \] 
        \STATE and update  the slope $m(t_k)$. 
    \ENDIF
\ENDFOR

\end{algorithmic}
\end{algorithm}

\subsection{Adaptation of the SBM for the MBL map}
\begin{proposition}
  The regularized objective function under the MBL map, $H_{(\varepsilon, \eta)}$ is a convex, differentiable and separable function \[ H_{(\varepsilon, \eta)}(\theta) = \sum_{i=1}^{n}  f_{\theta_i', c_i}^\eta(\theta_i),\]
  where $f_{\theta'_i, c_i}(x) = w_i \rho_{\eta_i}(\theta'_i- x) + \varepsilon w_i (x-c_i)^2,$ and the solution of $\partial f_{\theta_i', c_i}^\eta(x^*) =0$ exists and reads \[x^* = \begin{cases}
	\frac{\theta'_i + \varepsilon c_i}{(1+\varepsilon)} & \text{if } \mid \theta'_i-c_i \mid \leq \frac{1+\varepsilon}{\varepsilon} \eta_i \\
	c_i + \frac{\eta_i}{\varepsilon} & \text{if } \theta_i'-c_i > \frac{1+\varepsilon}{\varepsilon}\eta_i \\
	c_i - \frac{\eta_i}{\varepsilon} & \text{else}.
\end{cases}\]
\end{proposition}
\begin{proof}
  Immediate adaptation of Proposition~\ref{prop:certif_huber_sbm}'s
  proof with the derivative for each $i-$th coordinate
  being \[\partial f_{\theta_i', c_i}^\eta(x) = -2 w_i \min\bigg(1 \;
    , \;  \frac{\eta}{2w_i, \mid \theta_i'-x \mid}  \bigg) (\theta_i'-x)+ 2 \varepsilon w_i (x-c_i).  \]
\end{proof}
\noindent The function over the block $B_o$ reads 
\begin{align*}
	h_\eta(\theta) &= -2 \sum_{i \in B_o} w_i \min\!\left(1, \frac{\eta}{2w_i|\theta'_i - \theta|} \right)(\theta'_i - \theta) + 2\varepsilon \sum_{i \in B_o} w_i \, (\theta - T_\P(\theta')_i), \qquad \forall \theta \in \mathbb{R}.
\end{align*}
According to the induced weights \begin{equation} \label{eq:mbl_weights}
  w_{0, \eta_i} := \min\!\left(1; \dfrac{\eta }{2w_i|\theta'_i - \theta|} \right),
\end{equation} we define the following block sets 
\begin{equation*}
\begin{cases}
U(\theta)=\{i\in B_o \mid \theta<\theta'_i- \frac{\eta}{2w_i} \} & \text{(upper plateau)}\\
M(\theta)=\{i\in B_o \mid \theta'_i- \frac{\eta}{2w_i} \le\theta<\theta'_i+ \frac{\eta}{2w_i}\} & \text{(linear ramp)}\\
D(\theta)=\{i\in B_o \mid \theta'_i+ \frac{\eta}{2w_i} \le\theta\} & \text{(lower plateau)}
\end{cases}
\end{equation*}
thus the decomposition is, \begin{align*}
	h_\eta(\theta) = \underbrace{- \eta \bigg[\# U(\theta) - \# D(\theta)\bigg] -2 \sum_{j \in M(\theta)} w_j \theta'_j - 2 \varepsilon \sum_{j \in B_o} w_j T_\P(\theta')_j}_{=: \; c(\theta)} \; + \; \underbrace{2 \bigg[\sum_{j \in M(\theta)} w_j  + \varepsilon \sum_{j \in B_o} w_j\bigg]}_{=: \; m(\theta)} \, \theta
\end{align*}
The sets $U(\theta), M(\theta), D(\theta)$ only change at the $2\mid B_o \mid$ kink points
\[
\begin{cases}
	l_i = \theta_i' - \frac{\eta}{2w_i}, & i \in B_o \quad \text{(entry: $i$ moves $U \to M$)} \\
	L_i = \theta_i' + \frac{\eta}{2w_i}, & i \in B_o \quad \text{(exit: $i$ moves $M \to D$)}.
\end{cases}
\]
We apply and adapt the sweep Algorithm~\ref{alg:piecewise_root} to retrieve the exact solution of the MBL projection. 

\subsection{Adaptation of the SBM for the log-hyperbolic cosine map}

\begin{proposition}
Let $\kappa > 0$. The objective function associated with the projection map \eqref{eq:logcosh_map} is a convex, differentiable and separable function. Consequently, the Sequential Block Merging (SBM) algorithm converges to the unique global minimizer over $\P$.
\end{proposition}

\begin{proof}
By Proposition~\ref{prop:logcosh_map}, the objective function admits a unique minimizer over $\P$. Moreover, it satisfies Assumption 1 of \citet{wang_efficient_2022}, under which SBM is guaranteed to converge to an optimal solution. Combining these two results establishes convergence to the unique global minimizer.
\end{proof}

\noindent Applying SBM to this loss, requires solving the stationarity equation \eqref{eq:block_sol_logcosh} for an arbitrary block $B_o$. It is a transcendental equation and does not have an analytical expression. We use a Newton-Raphson method to retrieve the solution \begin{align*} 
	\theta_o^{(m+1)} &=  \theta_o^{(m)} - \frac{h_\kappa(\theta_o^{(m)})}{h'_\kappa(\theta_o^{(m)})}  \\
  &=  \theta_o^{(m)} +  \kappa \frac{\sum_{i \in B_o} w_i \tanh(\frac{\theta'_i - \theta_o^{(m)}}{\kappa})}{\sum_{i \in B_o} w_i sech^2(\frac{\theta'_i - \theta_o^{(m)}}{\kappa})} \quad \forall m \geq 0.
\end{align*}
Provided $h'_\kappa >0$, the convergence is quadratic, and the quadratic contraction constant is
\[
  \frac{h''_\kappa(\theta^\star)}{2\,h'_\kappa(\theta^\star)}
  = \frac{
      \displaystyle\sum_{i \in B_o}
        w_i\,\mathrm{sech}^2\!\left(\frac{\theta_i' - \theta^\star}{\kappa}\right)
        \tanh\!\left(\frac{\theta_i' - \theta^\star}{\kappa}\right)
    }{
      \kappa\displaystyle\sum_{i \in B_o}
        w_i\,\mathrm{sech}^2\!\left(\frac{\theta_i' - \theta^\star}{\kappa}\right)
    },
\]
hence,  the error $e^{(m)} = \theta^{(m)}- \theta^*$ satisfies \begin{equation*}
  e^{(m+1)} = \frac{h_\kappa''(\theta_o^*)}{2h'_\kappa(\theta_o^*)} e^{(m)^2} + O(e^{(m)^3}). 
\end{equation*}

\noindent The derivative is globally Lipschitz continuous, since \begin{equation*}
    \max_{x\in\mathbb{R}}
    \left|
        \operatorname{sech}^2(x)\tanh(x)
    \right|
    =
    \frac{2}{3\sqrt{3}},
\end{equation*} thus, the Lipschitz constant is 
$$  \frac{8}{3\sqrt{3}\kappa} \sum_{i\in B_o}w_i =:M_\kappa, \qquad \forall t\in\mathbb R. $$
Although \(M_\kappa\) is finite for every fixed \(\kappa>0\), it grows as \(1/\kappa\) as \(\kappa\) decreases. Consequently, for small values of \(\kappa\), the curvature of \(h_\kappa\) can vary rapidly, which may compromise the reliability of a pure Newton iteration. \\
The solution $\theta_o^*$ lies in the initial bracket
\[
  \theta_o^* \in [l^{(0)}, L^{(0)}], \qquad
  l^{(0)} := \min_{i\in B_o}\theta_i', \qquad
  L^{(0)} := \max_{i\in B_o}\theta_i'.
\]
The Newton iteration is safeguarded by maintaining this bracket $[l^{(m)}, L^{(m)}]$, which is updated after each bisection step and always satisfies $h_\kappa(l^{(m)}) \le 0 \le h_\kappa(L^{(m)})$.

\noindent Let $\theta_o^{(k)}$ be the current iterate, with Newton step
\[
  s^{(k)} = \frac{h_\kappa(\theta_o^{(k)})}{h_\kappa'(\theta_o^{(k)})}.
\]
We first propose the Newton candidate
\[
  \theta_o^{\mathrm{cand}} = \theta_o^{(k)} - s^{(k)}.
\]
When the Newton iteration is well behaved, the candidate remains inside the current bracket, $\theta_o^{\mathrm{cand}} \in [l^{(m)}, L^{(m)}]$, and we accept it: $\theta_o^{(k+1)} = \theta_o^{\mathrm{cand}}$.
Otherwise, the Newton step leaves the bracket and we perform a bisection step:
\[
  \theta_o^{(k+1)} = \frac{l^{(m)} + L^{(m)}}{2}.
\]

\noindent In both cases the bracket is then tightened using the sign of $h_\kappa$ at the new iterate:
\[
  \begin{cases}
    l^{(m+1)} = \theta_o^{(k+1)}, & \text{if } h_\kappa(\theta_o^{(k+1)}) < 0,\\[2mm]
    L^{(m+1)} = \theta_o^{(k+1)}, & \text{otherwise.}
  \end{cases}
\]

\noindent The iteration stops when the Newton iteration has converged,
\[
  |h_\kappa(\theta_o^{(k+1)})| \le \varepsilon_{\mathrm{match}}
  \quad \text{and} \quad
  |s^{(k+1)}| \le \varepsilon_{\mathrm{match}},
\]
or when bisection has reduced the bracket below the tolerance,
\[
  L^{(m)} - l^{(m)} < \varepsilon_{\mathrm{match}}.
\]
Hereafter, for numerical experiments, $\varepsilon_{\mathrm{match}}$ is set to the machine zero.

\section{ Posterior Projection } \label{sec:hyperparam}
The projection maps introduced above depend on their respective hyperparameters, namely \(\delta\), \(\eta\), and \(\kappa\). Propositions~\ref{prop:delta_threshold} and \ref{prop:eta_thereshold}, together with the heuristic established in~\eqref{eq:kappa_star}, identify ranges of plausible hyperparameter values for which the corresponding projections exhibit desirable properties. In practice, one must therefore select a unique hyperparameter for each posterior projection. Consider the projected posteriors associated to the three introduced maps \begin{align}
  \widetilde{\Pi}_\delta (\cdot \mid M) &= (T_\P^{(\delta)})_\# \Pi(\cdot \mid M) \label{eq:pushforward_huber} \\
  \widetilde{\Pi}_\eta (\cdot \mid M) &= (T_\P^{(\eta)})_\# \Pi(\cdot \mid M) \label{eq:pushforward_mbl}\\
  \widetilde{\Pi}_\kappa (\cdot \mid M) &= (T_\P^{(\kappa)})_\# \Pi(\cdot \mid M) \label{eq:pushforward_logcosh} 
\end{align}
Recall that the Huber and MBL maps refer to their regularized version. These three projected posterior distributions are well-defined thanks to their measurability. Indeed, $c=T_{\mathcal P}(\theta')$ is $1$-Lipschitz in $\theta'$ and the Huber and MBL objectives are $2\varepsilon$-strongly convex with parameters depending continuously on $\theta'$, the maps $\theta'\mapsto T^{(\delta)}_{\mathcal P}(\theta')$ and $\theta'\mapsto T^{(\eta)}_{\mathcal P}(\theta')$ are continuous. For the log-cosh map no regularization is needed: $H_\kappa$ is strictly convex (Proposition~\ref{prop:logcosh_map}), so its unique minimizer depends continuously on $\theta'$. 

\noindent In \cite{bouafia2026projectionbasedapproachbayesianage}, the pushforward distribution \eqref{eq:pushforward} associated to the quadratic objective function \eqref{eq:IR} admits a Wasserstein characterization. This follows from the results of \cite{astfalck_posterior_2024} regarding the metric projections that admits a unique solution. The three proposed maps are a projection only in the algebraic sense, their respective objective functions are not distances. 

\begin{figure}
  \centering
  \includegraphics[width = .8\linewidth]{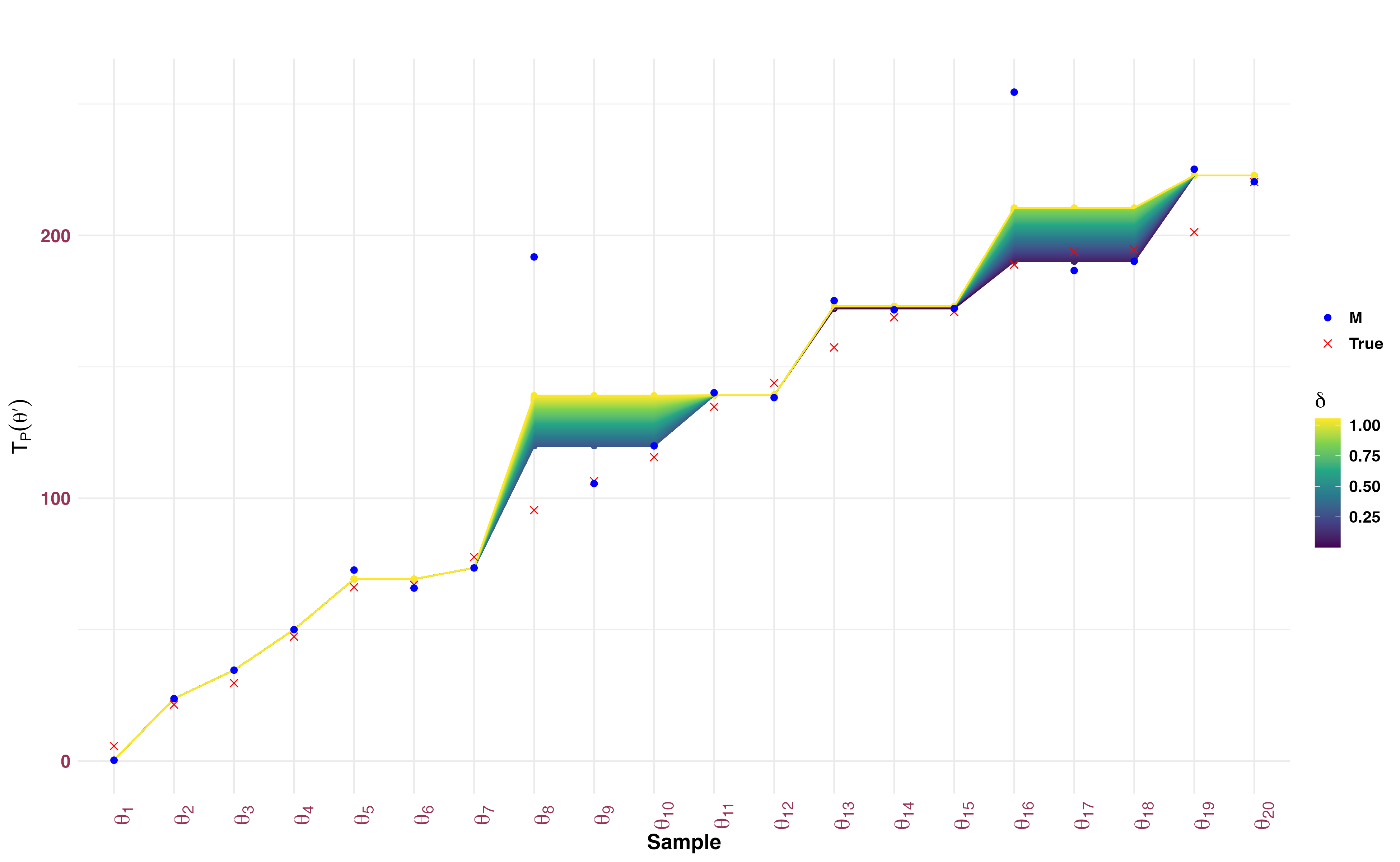}
  \caption{Solution paths $\bigl(\theta_{\eta}^*\bigr)_{\delta}$ under MBL $T_{\P}^{(\eta)}$ projections, over a grid of parameters $\eta$ under remote contamination at samples $\{8,16\}$ with a deviation of magnitude $a=80$, and a common weight value $w = 0.01$}
  \label{fig:mbl}
\end{figure}

\begin{figure}
  \centering
  \includegraphics[width = .7\linewidth]{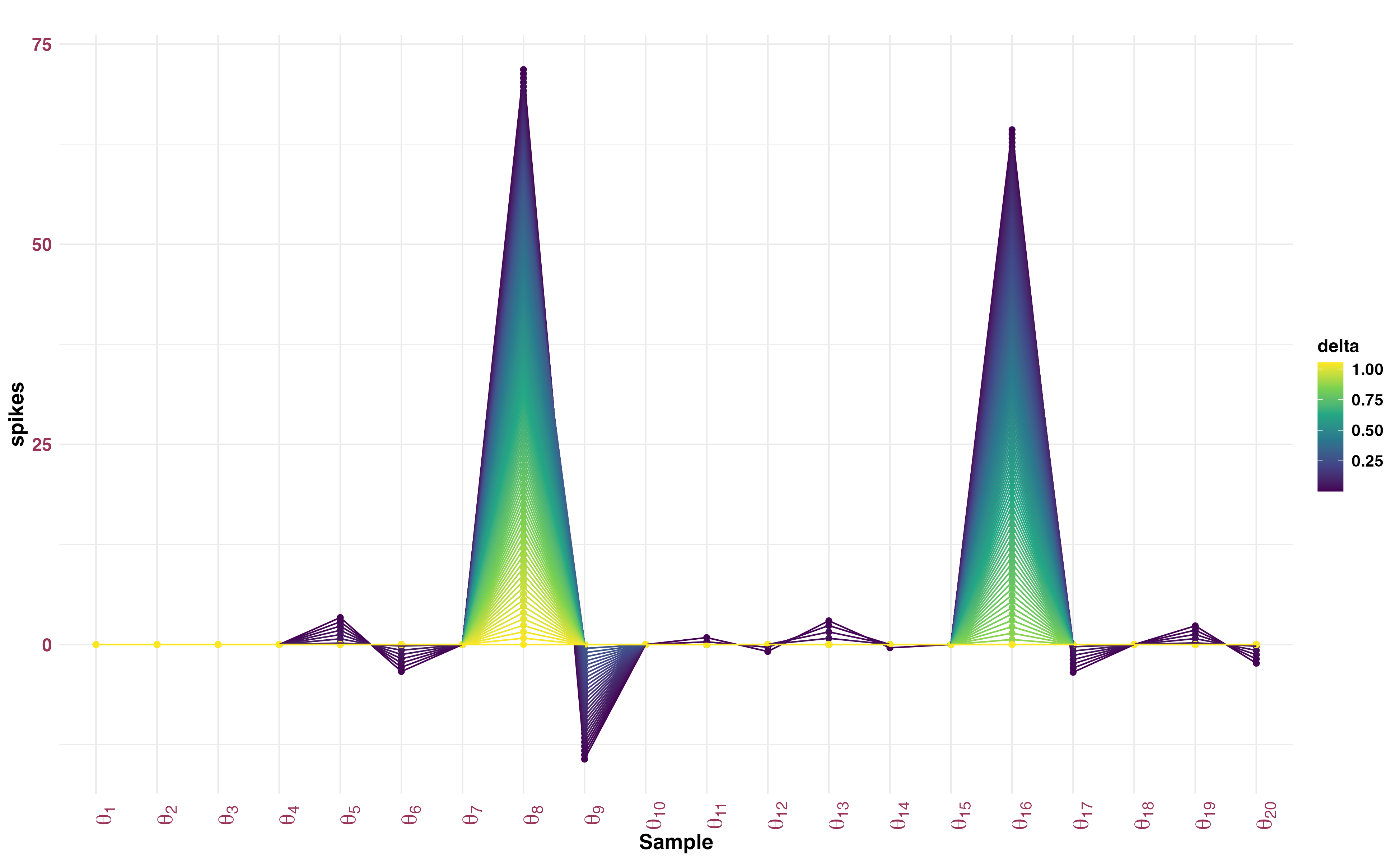}
  \caption{MBL spikes projection solution paths $\bigl(s_\eta^{*}\bigr)_{\eta}$ under remote contamination at samples $\{8,16\}$ with a deviation of magnitude $a=80$.}
  \label{fig:traj_spikes}
\end{figure}

\medskip
\noindent We propose a practical pipeline for selecting these hyperparameters. Hereafter, we fix the Tikhonov regularization term $\varepsilon = 10^{-7}$. The first step is to define a representative vector of the unconstrained posterior distribution \(\Pi(\cdot\mid M)\). We estimate its maximum a posteriori (MAP) estimate, defined as\[
\theta'_{\mathrm{MAP}}
=
\arg\max_{\xi\in\R^n}\pi(\xi\mid M),
\]
and its variances $\mathbb{V}ar(\theta_i \mid M_i)$, so we can define the corresponding weights $w_i$. 
We then construct a grid of candidate hyperparameter values for the Huber and MBL projection maps. Specifically, we consider\begin{align*}
\delta &\in  \left(
0,
\left\|
\theta'_{\mathrm{MAP}}-T_\P(\theta')
\right\|_\infty
\right] \\ 
\eta &\in
\left(
0,
\left\|
2W(\theta'_{\mathrm{MAP}}-T_\P(\theta'))
\right\|_\infty
\right].
\end{align*}
For each candidate value \(\delta_k\) and \(\eta_k\), we compute the corresponding projection of the MAP estimate:
\begin{align*}
  T_{\P}^{(\delta_k)}(\theta'_{\mathrm{MAP}}) \qquad T_{\P}^{(\eta_k)}
(\theta'_{\mathrm{MAP}}) \quad \forall k.
\end{align*}
This yields a family of solutions, which we refer to as projection trajectories, describing how the projected MAP estimate evolves as the hyperparameter varies.\\
The Figure~\ref{fig:mbl} displays the resulting trajectories for the MBL grids. For
this case study, we consider simulated observations $M\in\mathbb{R}^n$ generated according to the contamination scenario in \eqref{eq:contamination_scenario}. We set the deviation mean to $a=(80,80)$, the contaminated samples to $q_a=\{8,16\}$, and the common standard deviation to $\sigma=10$.
Assuming a flat prior, we fix the MAP $\theta'_{\mathrm{MAP}}$. The returned solution trajectories span from the Least Absolute Deviation (LAD) (values in purple) to the Isotonic regression solution (values in yellow). 
Further, a  spike component \(s^*\) can be visualized explicitly, for example in the case of the MBL, Figure~\ref{fig:traj_spikes} shows that the non-zero spikes are
concentrated around the contaminated samples \(q_a\). Their magnitude
increases as \(\eta\) decreases, reflecting the model's increasing
attribution of the discrepancy in the observations to the outlier
component rather than to the monotone baseline.
Note that the spike component is not specific to the MBL formulation (see Remark
\ref{rem:spike_existence}). 

\noindent The remaining question is how to choose the hyperparameter \(\lambda = \delta \text{ or }\eta\) of the posterior projections. A plausible approach is to use Stein's Unbiased Risk Estimate (SURE) \[ R = \norm{\theta - T_\P^{\lambda} (\theta')}_W^2, \]
where $\theta$ is the true value. 
The SURE requires evaluating the divergence of the projection map with respect to the observations $\frac{\partial}{\partial \theta'_i}T_\P^\lambda(\theta') $. For instance, in the Huber case, the resulting map is piecewise smooth: locally, once the active shape-constrained blocks and the Huber sets \begin{align*}
  B_o^+(\delta) &= \lbrace j \in B_o(\delta); \ \mid \theta'_j - \theta_o(\delta)\mid >\delta \rbrace \;\; B_o^-(\delta) = B_o(\delta) \setminus B_o^+(\delta) \quad  \forall o \in \lbrace 1, \cdots, K\rbrace, \\
  B_o^{+, \uparrow}(\delta) &= \{ j \in B_o^+; \ \theta'_j - \theta_o(\delta) > \delta\} \;\; B_o^{+, \downarrow}(\delta) = \{ j \in B_o^+; \ \theta'_j - \theta_o(\delta) < - \delta\}
\end{align*} are fixed, the estimator admits an explicit differentiable representation. This makes it possible to derive a candidate expression for its local divergence. In particular, in the unregularized case, the contribution of each active isotonic block is one, recovering the usual interpretation of the degrees of freedom as the number of fitted blocks $K$.
The difficulty is to turn this local calculation into a global almost-everywhere differentiability result. It is not sufficient to exclude only the points at which a residual lies exactly at the Huber transition, for example under the Huber map, the space would be \begin{equation*}
  \bar{\mathcal{G}} := \bigcup_{o=1}^K \bigcup_{j \in B_o}
  \Bigl\{\theta' \in \mathbb{R}^n :
    |\theta'_j - \theta_o(\theta')| = \delta  
  \Bigr\} =  \bigcup_{o=1}^K \bigcup_{j \in B_o} \bar{G}_{o, j}.
\end{equation*}
 One must also control the boundaries at which the active block structure changes, i.e. where adjacent constrained levels merge or split. A complete SURE argument would therefore require a careful characterization of the regions of the observation space on which 
\begin{enumerate}[(i)]
  \item the block structure $B_1 \cup \cdots \cup B_K$ for all $K \in \{1,\cdots, n\}$,
  \item the Huber $B_o^+, B_o^-$ pattern,
  \item the associated signs, i.e., $B_o^{+, \uparrow}, B_o^{+, \downarrow}$,
\end{enumerate}
remain fixed, together with a proof that their boundaries form a negligible set. This is closely related to the active-set arguments used for Huber regression with linear constraints in \cite{liu2021degrees}, and to the divergence approach for shape-restricted regression developed in \cite{meyer2000degrees}.
However, SURE is an unbiased estimates of the risk with respect to $\theta' \sim \Pi(. \mid M)$, which in the contamination scenario~\eqref{eq:contamination_scenario}, includes the contamination offsets, rather than with respect to the target chronology $\theta$. Minimizing it would therefore favour fitting the offsets, i.e. large values of $\lambda$, which defeats the purpose of the robust projection.

\noindent Instead, we use an empirical approach, based on the weights introduced in \eqref{eq:huber_weights} and \eqref{eq:mbl_weights}. These quantities provide an indication of the deviation of each unconstrained sample \(\theta'_i\) from the corresponding block estimate \(\theta_o\). Specifically, they separate the components into the quadratic part, \(B_o^-(\delta)\), whose contributions retain weights close to one, and the linear part, \(B_o^+(\delta)\), whose contributions are weighted by the inverse residual. For residuals larger than \(\delta\), this weight is smaller than one and decreases as the residual increases, making the contribution of strongly deviating observations increasingly negligible. Therefore, once a sensible value of \(\delta\) is reached, \(\delta\) is no longer artificially small, and only samples with large deviations remain above this threshold. We should then observe a marked change in the empirically decreasing function presented below.\\
For the Huber case, we set the decreasing function \begin{equation*}
  \psi(\delta) = \sum_{j=1}^{n} (1-w_{0,j}^{( \delta)}) =  n - \#( V^-(\delta)) - \delta \sum_{o=1}^{K} \sum_{j \in B_o^+(\delta)} (\mid \theta'_j - \theta_o (\delta) \mid)^{-1},
\end{equation*}
where $\#V^-(\delta) = \sum_{o = 1}^{K} \# B_o^-(\delta)$.
Alternatively, for the MBL case, the associated decreasing function is 
\begin{align*}
  \psi(\eta) &=  \sum_{j=1}^{n} (1-w_{0,\eta_j}) = n - \# V^-(\eta) - \eta \sum_{o =1}^{K} \sum_{j \in B_o^+} \frac{1}{2 w_j} (\mid \theta'_j - \theta_o(\eta) \mid )^{-1},
\end{align*}
where $\#V^-(\eta) = \sum_{o = 1}^{K} \# B_o^-(\eta)$. \\
The elbow of \(\psi\), marking the transition between the small- and large-\(\delta\) regimes, thus provides a separation between the noise and contamination scales. We use this feature to obtain a data-driven choice of \(\delta\), by applying an elbow detection method to the function \(\delta\mapsto\psi(\delta)\), evaluated on a regular grid of \(\delta\)-values. In the present numerical experiments, the grid has a length of $100$.
More specifically, we employ the Unit Invariant Knee (UIK) method introduced by \citet{christopoulos2016introducing}, which we apply to \(\psi\) to select the hyperparameters \(\delta^{\mathrm{UIK}}\) and \(\eta^\mathrm{UIK}\).
We use the heuristic~\eqref{eq:kappa_star}, to set the optimal value for the log-hyperbolic cosine map; \begin{equation*}
  \kappa^\mathrm{UIK} = \frac{\delta^\mathrm{UIK}}{\textrm{arctanh}(1-\nu)} \quad \nu >0 
\end{equation*}
The saturation parameter is set to $\nu = 0.04$.  

\section{Simulation} \label{sec:simulation}
In this section, we consider \(p\) independent observations
\((M^{(j)})_{j=1}^p\), generated according to the contamination
scenario described in Equation~\eqref{eq:contamination_scenario},
under four different configurations of the contamination parameters:
\begin{enumerate}
    \item \textbf{Use case 1 (uncontaminated scenario):}
    We consider a reference setting with no contamination, that is,
    \(q_a = \emptyset\).

    \item \textbf{Use case 2 (moderate contamination):}
    We contaminate two out of the 20 samples, with a common deviation
    mean \(a = (60,60)\) and a common standard deviation
    \(\sigma_i = \sigma = 20\) for all samples. We consider
    two configurations:
    \[
    q_a =
    \begin{cases}
        \{7,8\}, & \text{consecutive contamination},\\
        \{8,16\}, & \text{distant contamination}.
    \end{cases}
    \]

    \item \textbf{Use case 3 (strong contamination):}
    We again contaminate two out of the 20 samples, but with a larger
    deviation mean \(a = (80,80)\). The standard deviation remains
    common to all contaminated samples, with
    \(\sigma_i = \sigma = 10\). As in the previous setting, we consider
    two configurations:
    \[
    q_a =
    \begin{cases}
        \{7,8\}, & \text{consecutive contamination},\\
        \{8,16\}, & \text{distant contamination}.
    \end{cases}
    \]

    \item \textbf{Use case 4 (multiple contaminations):}
    We consider a more complex scenario in which five out of the
    20 samples are contaminated, with different deviation means
    \(a = (20,80,-50,-80,60)\). The standard deviations
    vary across samples such as 
    \(\sigma_i \in [6,20]\). The contaminated samples are located at
    \(q_a = \{7,8,11,15,18\}\), combining consecutive and distant
    contaminations.
\end{enumerate}

\begin{figure}
  \centering
  \includegraphics[width = \linewidth]{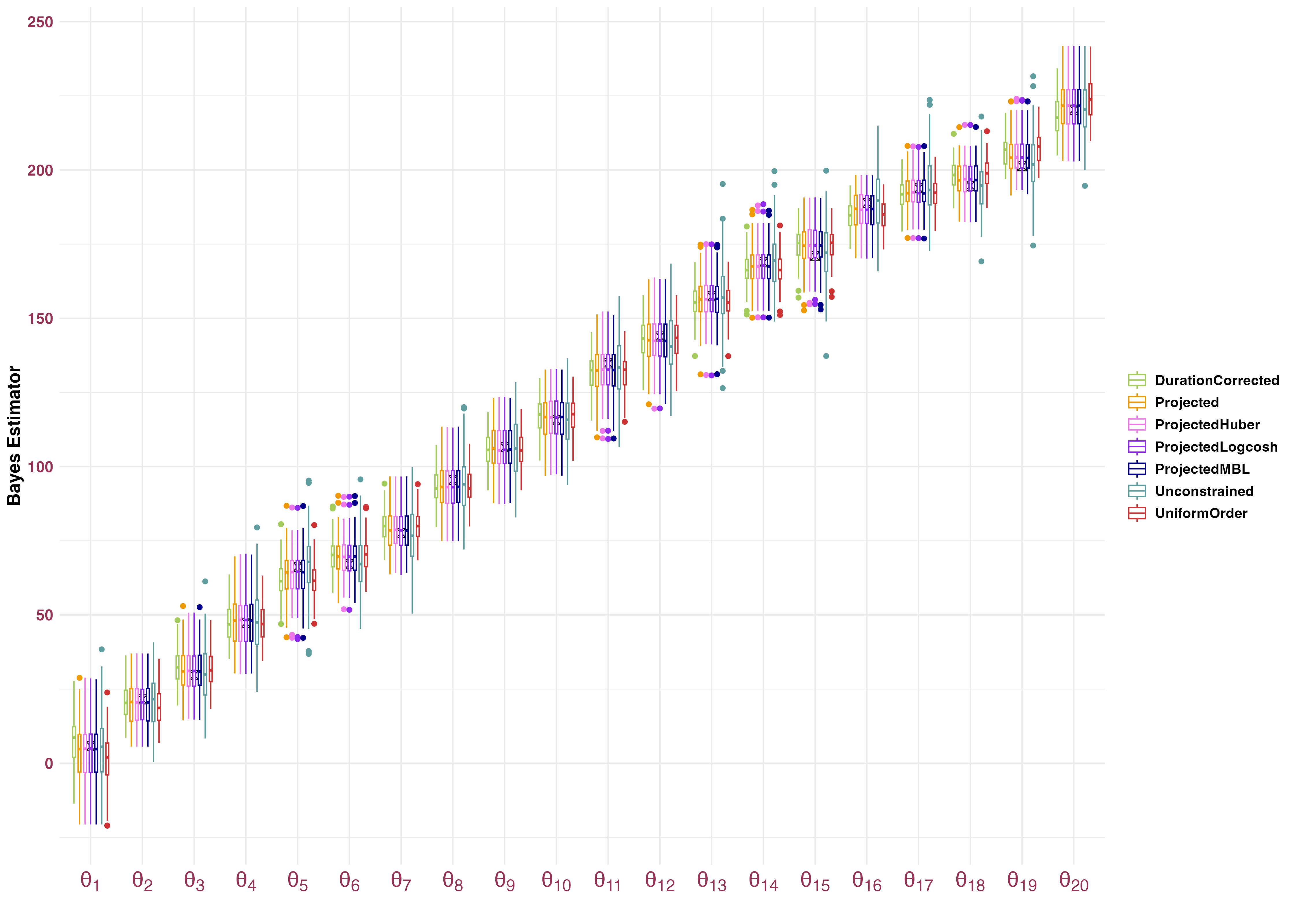}
  \caption{Boxplots of the Bayes estimators obtained from 100 replications of the simulated observations \(M\) under Use case 1 (uncontaminated scenario). Black square markers indicate the true age values \(\theta\).}
  \label{fig:no_contamination}
\end{figure}

\noindent For each use case and observation \(M^{(j)}\), we determine the optimal hyperparameters using the procedures described in Section~\ref{sec:hyperparam}. We then compare the resulting posterior distributions under each of the considered models:
\begin{itemize}
\item  Reference model 
\begin{itemize}
    \item Unconstrained model:
      $\Pi(\cdot\mid M)$ obtained with a flat prior, defined in \eqref{eq:flat}.
\end{itemize}

\item Classical Bayesian approaches,  see \cite{bouafia2026projectionbasedapproachbayesianage} for more details
\begin{itemize}    
    \item Uniform-order model:
          $\Pi_{\mathcal P}(\cdot\mid M)$ using the classical uniform-order prior \eqref{eq:uodensity}.
    
    \item Corrected uniform-order model (Duration Corrected):
          $\Pi_{\mathcal P}^{N}(\cdot\mid M)$ using the duration correction proposed by \citet{nicholls_radiocarbon_2001}.
\end{itemize}

\item Projected-posterior
\begin{itemize}
    \item Projected $\widetilde{\Pi}(\cdot\mid M)$, defined in \eqref{eq:pushforward}
    \item Projected Huber $\widetilde{\Pi}_{\delta^\mathrm{UIK}}(\cdot\mid M)$, defined in \eqref{eq:pushforward_huber}
    \item  Projected MBL $\widetilde{\Pi}_{\eta^\mathrm{UIK}}(\cdot\mid M)$, defined in \eqref{eq:pushforward_mbl}
    \item Projected Logcosh $\widetilde{\Pi}_{\kappa^\mathrm{UIK}}(\cdot\mid M)$, defined in \eqref{eq:pushforward_logcosh}
\end{itemize}
\end{itemize}
Figure~\ref{fig:no_contamination} displays the Bayes estimators obtained for each model under Use case 1. Overall, the results are satisfactory, with all robust models achieving performance comparable to that of the standard Projected posterior. As expected, they exhibit some loss of precision compared with the non-robust models. Nevertheless, their accuracy remains satisfactory, although it is somewhat lower for certain samples, particularly for the Uniform Order posterior. This is notably the case for sample 20, a behavior that is consistent with the duration inflation issues inherent to this model (see \cite{bouafia2026projectionbasedapproachbayesianage} for further details).

\begin{figure}
  \centering
  \includegraphics[width = .95\linewidth]{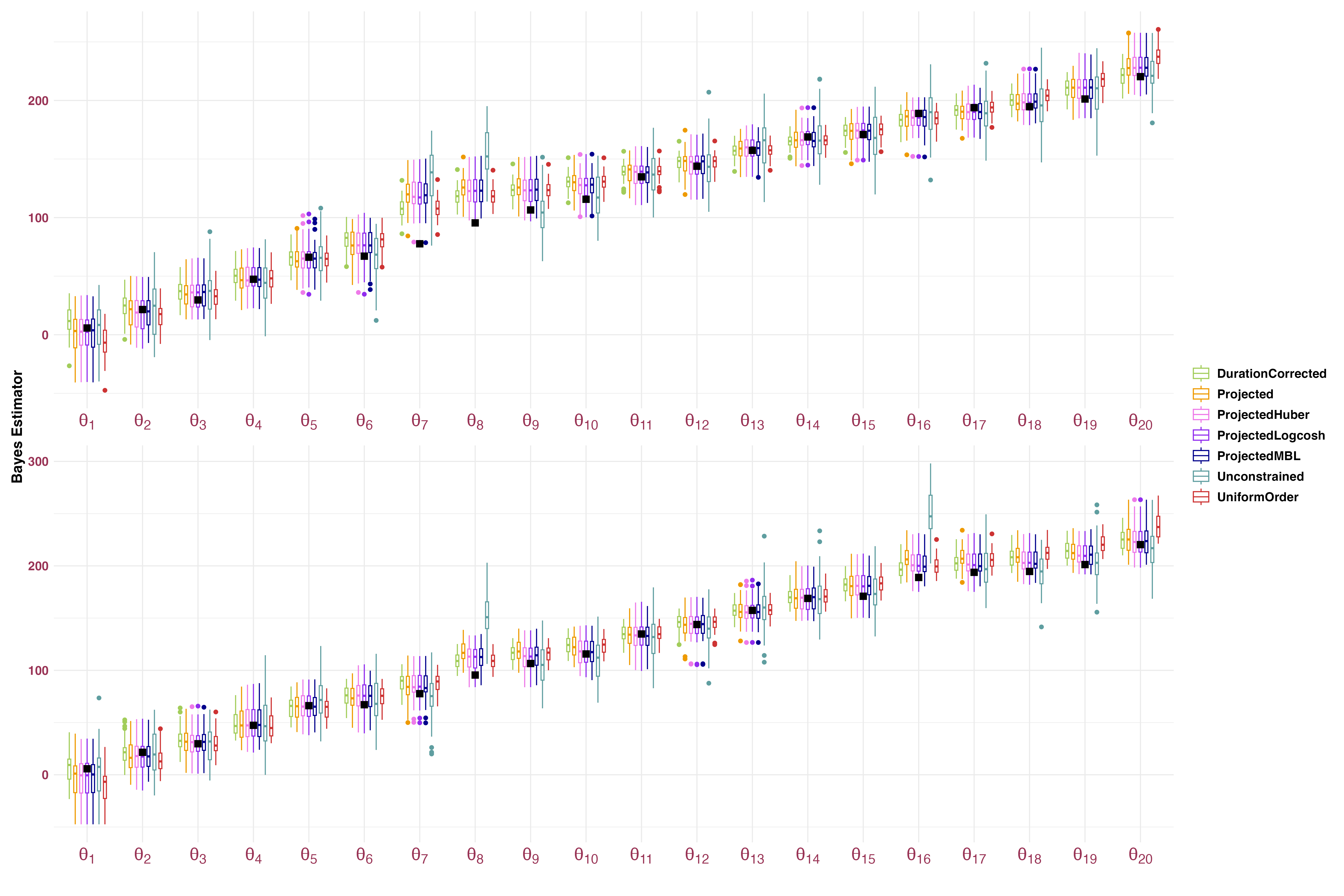}
  \caption{Boxplots of the Bayes estimators obtained from 100 replications of the simulated observations \(M\) under Use case 2 (moderate contamination), with \(q_a = \{7,8\}\) (top) and \(q_a = \{8,16\}\) (bottom). Black square markers indicate the true age values \(\theta\).}
  \label{fig:pos_M60_sd20}
\end{figure}

\begin{figure}
  \centering
  \includegraphics[width = .95\linewidth]{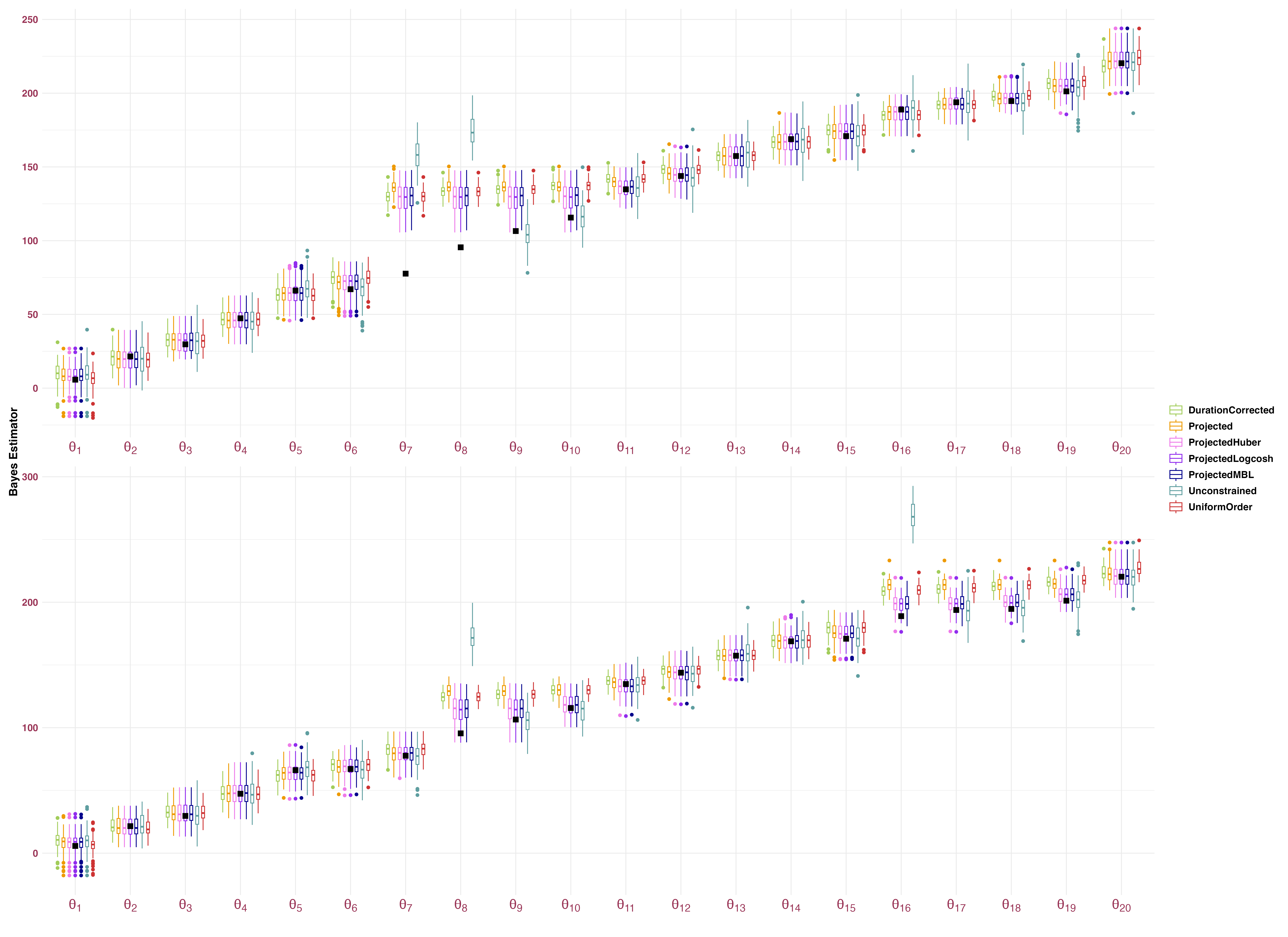}
  \caption{Boxplots of the Bayes estimators obtained from 100 replications of the simulated observations \(M\) under Use case 3 (strong contamination), with \(q_a = \{7,8\}\) (top) and \(q_a = \{8,16\}\) (bottom). Black square markers indicate the true age values \(\theta\).}
  \label{fig:pos_M80_sd10}
\end{figure}

\noindent  Figure~\ref{fig:pos_M60_sd20} displays Bayes estimators under Use case 2. Under the consecutive contamination configuration, the robust projection provides a modest improvement over the quadratic projection, particularly in the neighborhood of the contaminated samples. 
Under the distant contamination , the overall posterior estimates are satisfactory for both approaches, with the robust projection providing a slight improvement in the vicinity of the contaminated samples.
 Nevertheless, Use case 2 is relatively favorable: the contamination deviations are not sufficiently large to produce substantial separation from the uncontaminated measurements. To investigate a more challenging setting, we now consider Use case 3, which involves less noisy measurements and a larger mean deviation. Figure~\ref{fig:pos_M80_sd10} shows the Bayes estimators obtained for each model. Both contamination configurations highlight the improvements brought by the robust projected posterior.

\noindent Tables~\ref{tab:coverage_posM80} and \ref{tab:mse_posM80} report the coverage and MSE results for the consecutive contamination configuration, \(q_a = \{7,8\}\), and its neighboring samples. The uncontaminated neighboring samples, namely samples \(9\) and \(10\), exhibit lower MSE values and non-null coverage under the robust projected posterior, unlike under the non-robust models.

\noindent Note that the Unconstrained posterior is not reported in the tables, as its performance is not of interest in the present analysis.
\begin{table}
\centering
\small
\begin{tabular}{lcccccc}
\toprule
Sample & DurationCorrected & Projected & ProjectedHuber & ProjectedLogcosh & ProjectedMBL & UniformOrder \\
\midrule
6 & 0.95 & \cellcolor{green!25}\textbf{0.97} & 0.96 & 0.96 & \cellcolor{green!25}\textbf{0.97} & 0.94 \\
7 & 0.00 & 0.00 & 0.00 & 0.00 & 0.00 & 0.00 \\
8 & 0.00 & 0.00 & 0.00 & 0.00 & 0.00 & 0.00 \\
9 & 0.00 & 0.00 & 0.09 & \cellcolor{green!25}\textbf{0.11} & 0.00 & 0.00 \\
10 & 0.01 & 0.01 & 0.47 & \cellcolor{green!25}\textbf{0.53} & 0.02 & 0.01 \\
11 & 0.73 & 0.92 & \cellcolor{green!25}\textbf{0.99} & \cellcolor{green!25}\textbf{0.99} & 0.92 & 0.72 \\
12 & 0.95 & 0.95 & 0.95 & \cellcolor{green!25}\textbf{0.96} & 0.95 & 0.95 \\
\bottomrule
\end{tabular}
\caption{Coverage of the 95\% credible intervals of the projected posteriors over 100 replications under Use case 3 (strong contamination), with \(q_a = \{7,8\}\).  }
\label{tab:coverage_posM80}
\end{table}

\begin{table}
\centering
\small
\begin{tabular}{lcccccc}
\toprule
Sample & DurationCorrected & Projected & ProjectedHuber & ProjectedLogcosh & ProjectedMBL & UniformOrder \\
\midrule
6 & 93.72 & 69.80 & 77.10 & 76.91 & \cellcolor{green!25}\textbf{69.80} & 93.78 \\
7 & \cellcolor{green!25}\textbf{2756.12} & 3470.03 & 2827.14 & 2766.58 & 3424.90 & 2758.30 \\
8 & 1464.52 & 1714.11 & 1266.00 & \cellcolor{green!25}\textbf{1226.06} & 1677.54 & 1465.63 \\
9 & 825.63 & 927.91 & 617.19 & \cellcolor{green!25}\textbf{590.21} & 901.51 & 826.65 \\
10 & 489.25 & 463.64 & 264.87 & \cellcolor{green!25}\textbf{248.48} & 444.70 & 489.57 \\
11 & 67.93 & 46.42 & \cellcolor{green!25}\textbf{40.92} & 42.61 & 44.47 & 68.27 \\
12 & 38.88 & 40.79 & 47.04 & 47.80 & 41.61 & \cellcolor{green!25}\textbf{38.77} \\
\bottomrule
\end{tabular}
\caption{Mean Squared Error (MSE) of the Bayes estimators over 100 replications under Use case 3 (strong contamination), with \(q_a = \{7,8\}\). }
\label{tab:mse_posM80}
\end{table}

\begin{figure}
  \centering
  \includegraphics[width = \linewidth]{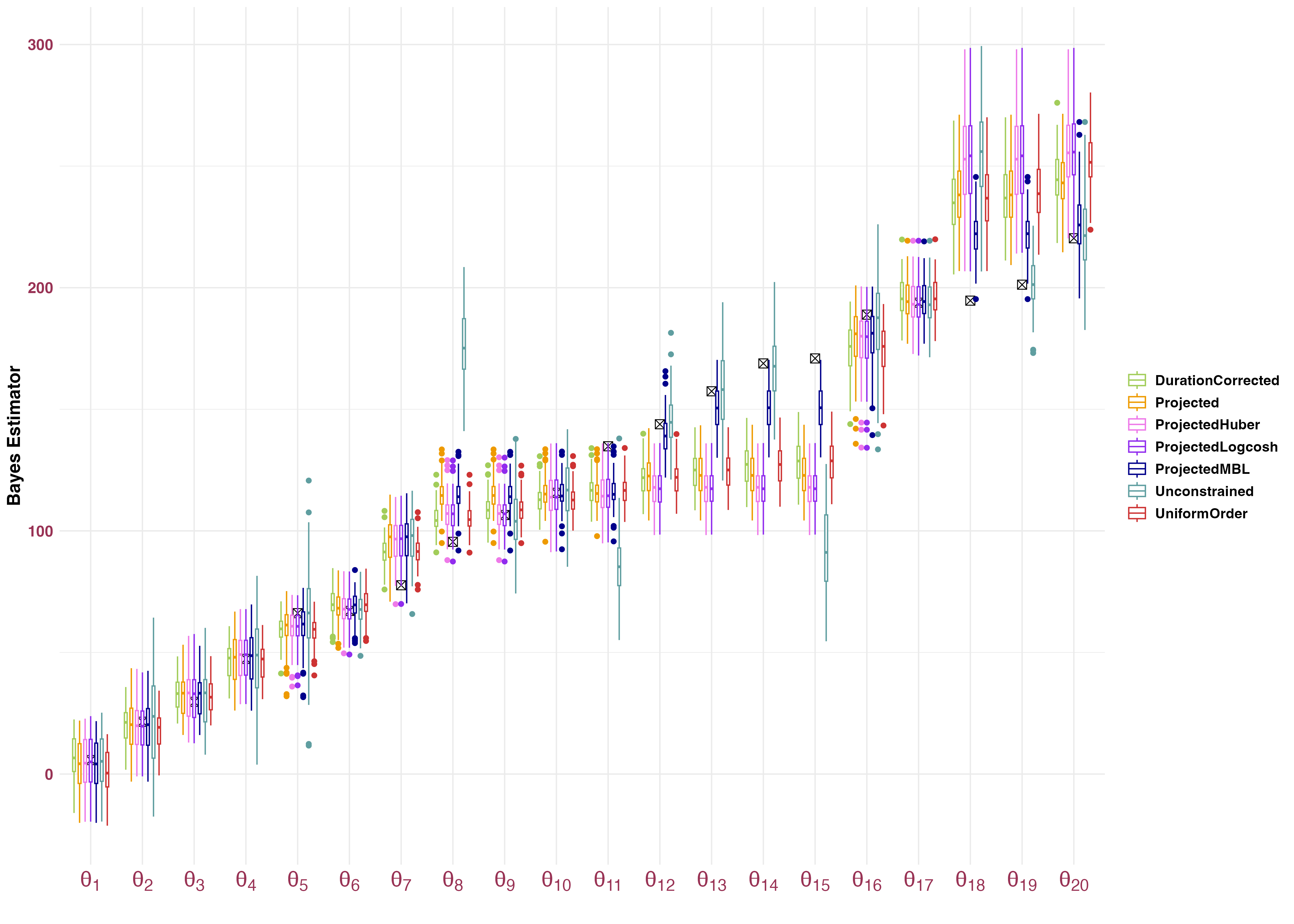}
  \caption{Boxplots of the Bayes estimators obtained from 100 replications of the simulated observations \(M\) under Use case 4 (multiple contamination), with \(q_a = \{7,8, 11,15,18\}\). Black square markers indicate the true age values \(\theta\).}
  \label{fig:meanboxplot_exotic}
\end{figure}

\noindent Finally, in Tables~\ref{tab:coverage_hybrid_osc}, \ref{tab:mse_hybrid_osc}, and Figure~\ref{fig:meanboxplot_exotic}, we consider Use case 4. In Tables, we report results only for the contaminated observations and their neighboring samples, as well as for the samples most affected by contamination. 

\noindent The non-robust models, Duration Corrected and Uniform Order posteriors, perform better around samples 7 and 8. Their neighboring samples, which may also be affected by contamination (here, sample 9), are not substantially impacted, as reflected by their satisfactory coverage and MSE results. At the next contaminated sample (sample 11), the non-robust models exhibit poorer coverage than ProjectedHuber and ProjectedLogcosh. However, they achieve smaller MSE values, illustrating the trade-off between variance and bias: using the Huber and Logcosh maps entails some loss of precision in exchange for greater accuracy.

\noindent For samples 12 to 20, the results are markedly different. With the exception of samples 16 and 17, which are not directly affected by any contamination, all models fail to recover the true underlying structure, resulting in null coverage and substantially larger MSE values. The good performance observed at samples 16 and 17 suggests that, in the absence of contamination, all models perform well when the underlying unconstrained models exhibit limited overlap between samples. The only exception to this overall deterioration is ProjectedMBL, which consistently provides satisfactory coverage and the smallest MSE values. Its weakest coverage occurs at sample 18, where contamination is present.

\noindent At first glance, the good performance of the non-robust models at the beginning of this section may seem surprising. However, understanding this behavior requires considering the entire structure rather than examining each sample independently. Indeed, the non-robust models are constructed to preserve monotonicity by progressively increasing the estimated values along the underlying order structure, which in this case is a chain. Consequently, contamination at sample 15 slows down this progression, leading these models to assign smaller values to subsequent samples than would be expected based on the uncontaminated observations at samples 12--14.

\begin{table}
\centering
\small
\begin{tabular}{lcccccc}
\toprule
Sample & DurationCorrected & Projected & ProjectedHuber & ProjectedLogcosh & ProjectedMBL & UniformOrder \\
\midrule
7 & \cellcolor{green!25}\textbf{0.76} & 0.63 & 0.67 & 0.67 & 0.67 & \cellcolor{green!25}\textbf{0.76} \\
8 & 0.81 & 0.11 & 0.88 & \cellcolor{green!25}\textbf{0.89} & 0.31 & 0.81 \\
9 & 0.95 & 0.75 & 0.96 & \cellcolor{green!25}\textbf{0.97} & 0.90 & 0.95 \\
11 & 0.14 & 0.11 & \cellcolor{green!25}\textbf{0.48} & \cellcolor{green!25}\textbf{0.48} & 0.28 & 0.14 \\
12 & 0.10 & 0.19 & 0.12 & 0.16 & \cellcolor{green!25}\textbf{0.98} & 0.10 \\
13 & 0.00 & 0.00 & 0.00 & 0.00 & \cellcolor{green!25}\textbf{0.91} & 0.00 \\
14 & 0.00 & 0.00 & 0.00 & 0.00 & \cellcolor{green!25}\textbf{0.65} & 0.00 \\
15 & 0.00 & 0.00 & 0.00 & 0.00 & \cellcolor{green!25}\textbf{0.62} & 0.00 \\
16 & 0.83 & \cellcolor{green!25}\textbf{0.94} & \cellcolor{green!25}\textbf{0.94} & 0.93 & 0.93 & 0.83 \\
17 & \cellcolor{green!25}\textbf{0.97} & \cellcolor{green!25}\textbf{0.97} & 0.96 & 0.96 & 0.96 & \cellcolor{green!25}\textbf{0.97} \\
18 & 0.00 & 0.00 & 0.01 & 0.01 & \cellcolor{green!25}\textbf{0.26} & 0.00 \\
19 & 0.02 & 0.04 & 0.01 & 0.01 & \cellcolor{green!25}\textbf{0.61} & 0.01 \\
20 & 0.22 & 0.34 & 0.10 & 0.08 & \cellcolor{green!25}\textbf{0.97} & 0.22 \\
\bottomrule
\end{tabular}
\caption{Coverage of the 95\% credible intervals of the projected
  posteriors over 100 replications under a more complex contamination
  scenario, with $q_a = \{7,8,11,15,18\}$ and contamination means $a =
  (20,80,-50,-80,60)$, for different standard deviations $\sigma_i $
  are   uniformly sampled  on $[6,20]$.} 
\label{tab:coverage_hybrid_osc}
\end{table}

\begin{table}
\centering
\small
\begin{tabular}{lcccccc}
\toprule
Sample & DurationCorrected & Projected & ProjectedHuber & ProjectedLogcosh & ProjectedMBL & UniformOrder \\
\midrule
7 & \cellcolor{green!25}\textbf{217.35} & 427.55 & 410.97 & 418.21 & 440.42 & 219.20 \\
8 & \cellcolor{green!25}\textbf{116.55} & 413.67 & 176.99 & 174.51 & 393.44 & 117.50 \\
9 & 35.71 & 108.96 & 49.96 & 50.70 & 103.43 & \cellcolor{green!25}\textbf{35.59} \\
11 & 374.15 & 392.70 & 455.54 & 447.21 & 393.74 & \cellcolor{green!25}\textbf{373.49} \\
12 & 538.07 & 529.84 & 751.59 & 748.24 & \cellcolor{green!25}\textbf{93.68} & 535.76 \\
13 & 1115.55 & 1259.43 & 1638.62 & 1634.81 & \cellcolor{green!25}\textbf{140.91} & 1112.30 \\
14 & 1805.53 & 2171.47 & 2677.12 & 2672.39 & \cellcolor{green!25}\textbf{429.04} & 1801.38 \\
15 & 1853.13 & 2363.60 & 2891.57 & 2886.65 & \cellcolor{green!25}\textbf{508.87} & 1848.87 \\
16 & 312.38 & 267.69 & 282.28 & 284.18 & \cellcolor{green!25}\textbf{228.77} & 314.55 \\
17 & \cellcolor{green!25}\textbf{74.04} & 77.38 & 86.42 & 87.64 & 75.51 & 74.89 \\
18 & 1825.25 & 2085.24 & 3754.70 & 3853.32 & \cellcolor{green!25}\textbf{805.62} & 1967.00 \\
19 & 1486.90 & 1559.82 & 3039.59 & 3127.35 & \cellcolor{green!25}\textbf{496.84} & 1630.56 \\
20 & 745.49 & 683.99 & 1567.56 & 1620.71 & \cellcolor{green!25}\textbf{204.99} & 1109.37 \\
\bottomrule
\end{tabular}
\caption{Mean squared error (MSE) of the Bayes estimators over 100
  replications under a more complex contamination scenario, with $q_a
  = \{7,8,11,15,18\}$ and contamination means $a =
  (20,80,-50,-80,60)$, for different standard deviations   are
  uniformly sampled on $[6,20]$. } 
\label{tab:mse_hybrid_osc}
\end{table}

\section{Application to geochronological case studies} \label{sec:real_applications}

\subsection{Loess sequences as climatic archives}
Loess sequences consist of sediment made of mineral dust accumulated by wind over long periods of time, often interspersed by ancient soils. Such deposits are important climatic archives, because loess deposition requires open environments and strong winds, generally associated with global glacial periods, while soil formation implies vegetation growth and humid conditions (\cite{frechen2003loess,stevens2013genetic}). Therefore, obtaining precise and robust chronologies is important to reconstruct climatic and environmental fluctuations in time (\citet{lu2007new,thiel2011luminescence}). From a stratigraphic perspective, loess-palaeo-soil sequences form rather simple case studies, because altitude is a direct proxy for age - more elevated samples being systematically younger than lower ones.
\subsubsection{Jingbian site (D3 section)}

The first application uses the Jingbian dataset published by \cite{stevens_ice-volume-forced_2018}, which consists of sediment deposits from the Chinese Loess Plateau in Jingbian, China. The dataset selected here contains 54 Optically Stimulated Luminescence (OSL) ages and samples are supposed to follow a chain order structure (total ordering). For this application, we use the Bayesian model introduced by \cite{combes_bayesian_2017} as the unconstrained model.

\noindent This dataset presents two challenges for chronological modeling. First, some sections contain samples with statistically indistinguishable measurements. Given the limited precision of OSL dating, these measurements can result in apparent age inversions, even when the underlying stratigraphic ordering is preserved. Second, samples 25--31 form a section in which samples 30 and 31 exhibit large deviations from the expected sequence. This provides a setting in which the influence of individual deviating observations can be assessed using the robust projection methods.

\begin{figure}
  \centering
  \includegraphics[width= \linewidth]{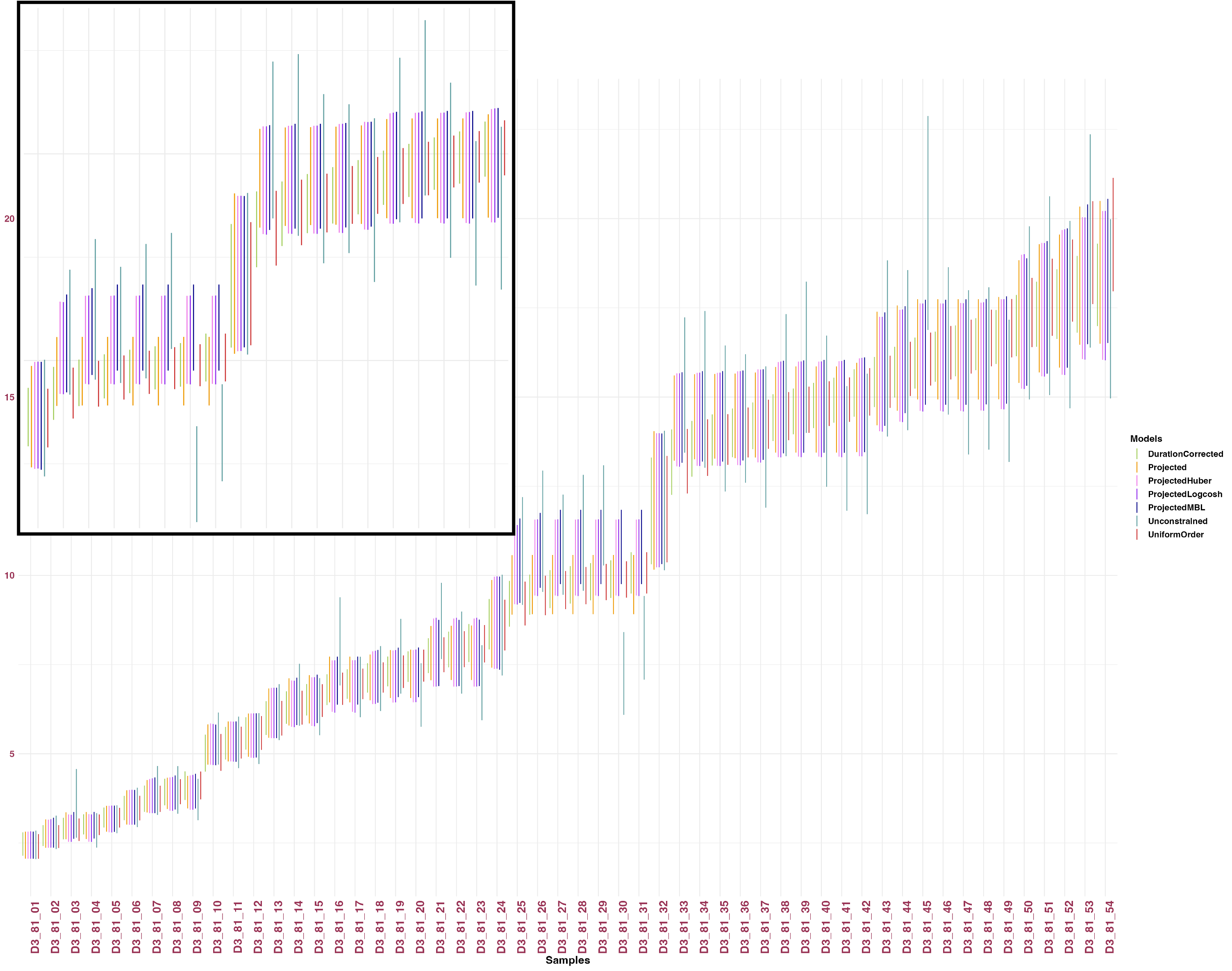}
  \caption{Comparison of the 95\% credible intervals obtained with the Unconstrained, Projected, Projected Log-hyperbolic cosine, Projected MBL at the Jingbian site.}
  \label{fig:hpd_jingbian}
\end{figure}

\noindent Figure~\ref{fig:hpd_jingbian} illustrates the differences
between the projection methods. The non-robust projections are
influenced by the deviating samples, resulting in younger estimated
ages across the affected section. In contrast, the robust projections
reduce the influence of these observations and yield older age
estimates for the section. This effect is slightly more pronounced for the MBL projected posterior.

\subsubsection{Xifeng site}
For the second application, we consider the Xifeng dataset from \cite{stevens2016mass}, which consists of sediment deposits from the Loess Plateau at Xifeng, China. The dataset comprises 49 OSL samples, for which we impose a chain-order structure as the stratigraphic constraint. As in the Jingbian dataset, the limited precision of the measurements leads to numerous apparent inversions in the estimated ages.

\begin{figure}
  \centering
  \includegraphics[width= \linewidth]{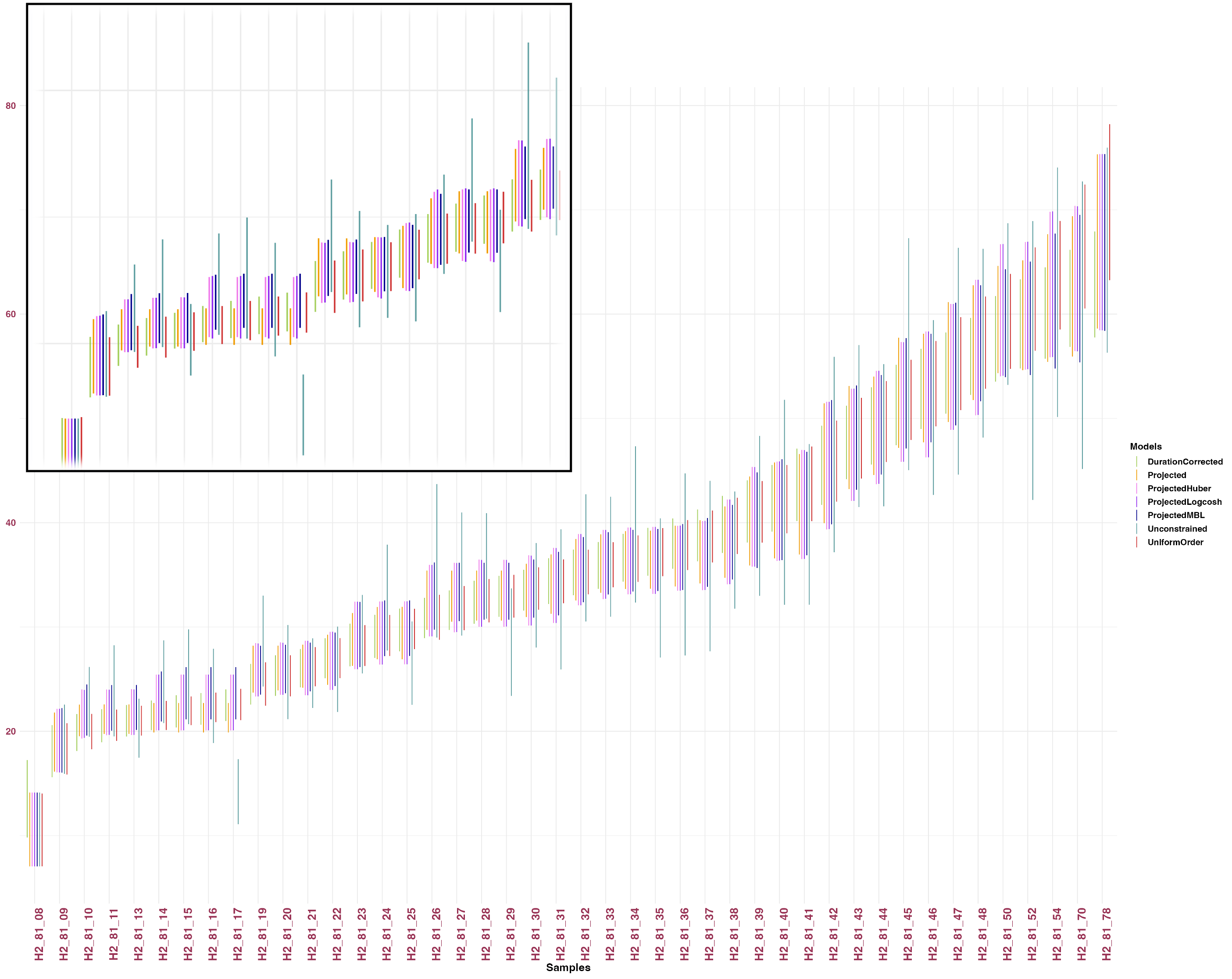}
  \caption{Comparison of the 95\% credible intervals obtained with the Unconstrained, Classical Bayesian Approach, Projected, Projected Log-hyperbolic cosine and Projected MBL for OSL measures at Xifeng Site.}
  \label{fig:hpd_xifeng}
\end{figure}

\noindent The main challenge for the robust methods occurs at the beginning of the section (see Figure~\ref{fig:hpd_xifeng}), where two deviating samples, one with a moderate deviation and the other with a larger deviation, affect the age estimates over samples 10--21. The non-robust projections tend to produce younger age estimates for the non-deviating samples in this section. The robust projections reduce the influence of these deviations, with the MBL projected posterior showing a more pronounced effect, particularly in terms of the precision of the resulting estimates.

In both case studies, the loess sections of Jingbian and Xifeng, the three tested robust models (ProjectedHuber, ProjectedLogcosh and ProjectedMBL) provide significant improvements in terms of robustness. More specifically, it seems that the model least affected by samples appearing as outliers is the MBL.

\subsection{Archaeological sequences}
In archaeology, stratigraphies are generally less simple for several
reasons: (i) the sedimentary deposits are split in units that are often not delimited by flat, horizontal surfaces (for example, because a pit was dug in the soil, or because cave deposits are not formed on flat bedrocks and rather follow the geological morphology); (ii) unlike loess deposits accumulated over wide surfaces, these deposits vary locally, from one part of an archaeological site to another; (iii) in general, several samples are taken from each stratigraphic unit, inside of which ordering constraints cannot be determined. As a result, the graphs corresponding to the stratigraphic constraints are more complex.     

\subsubsection{The Middle Palaeolithic site of Roc-en-Pail}
The third application concerns a Middle Palaeolithic site located in
Chalonnes-sur-Loire (France), which was recently excavated by Sylvain
Soriano and his team (unpublished). In such case studies, numerical chronology mainly serves two purposes: first, to compare human occupations from distant sites and/or regions; second, to place the human occupations on a time scale to allow discussing their modalities (lithic industries, subsistence strategies, site function, degree of fire use, etc.) in light of their climatic and environmental conditions. For both aims, large uncertainties hamper insightful comparisons and contextualization.

\begin{figure}
  \centering
  \includegraphics[width= \linewidth]{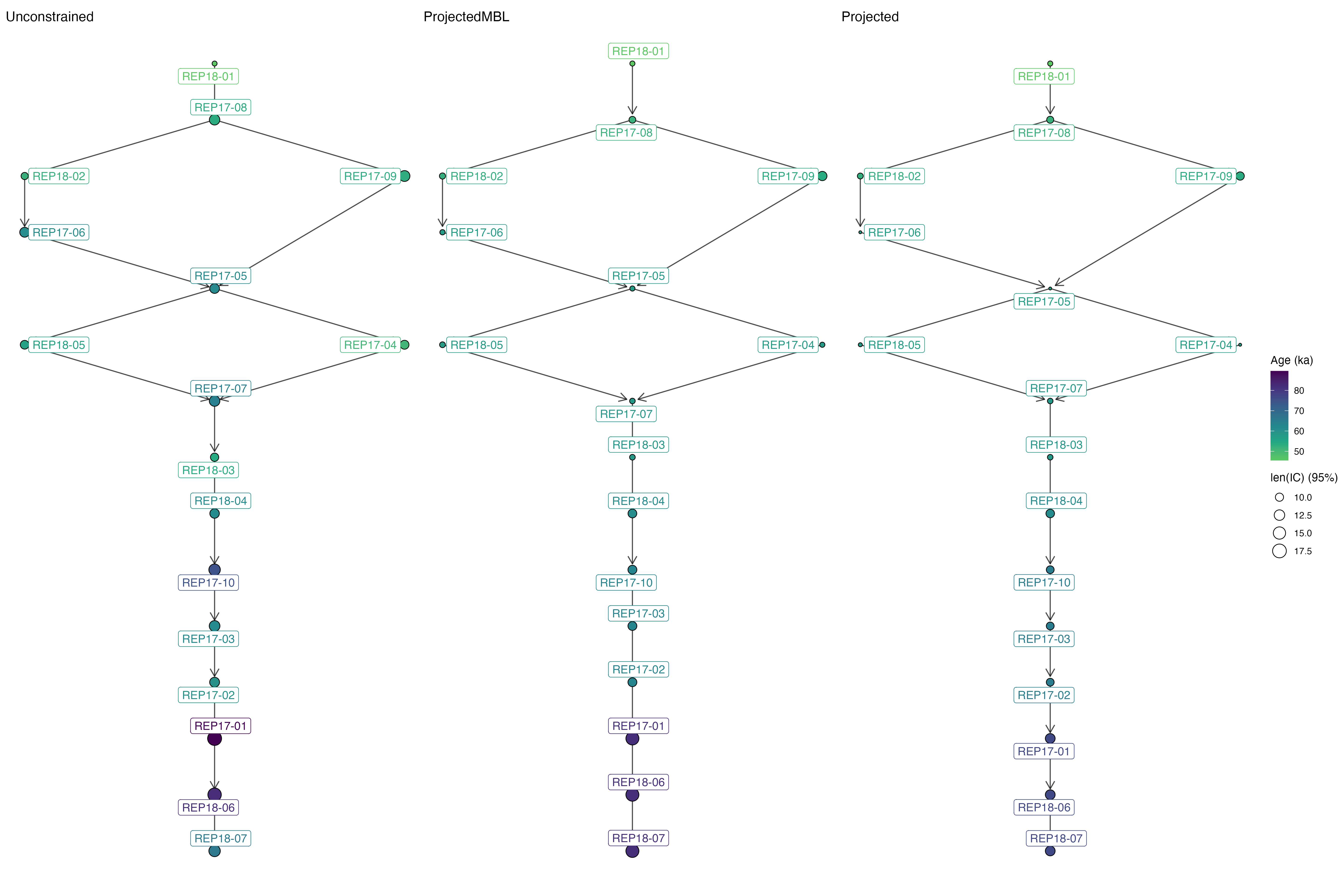}
  \caption{Directed Acyclic Graph (DAG) of the stratigraphic constraints for the Roc-en-Pail datasets. Node colors represent the mean ages and node sizes are proportional to the length of the corresponding 95\% credible intervals of each posterior, Unconstrained, Projected MBL and Projected.}
  \label{fig:dag_rocenpail}
\end{figure}

\noindent The dataset comprises 17 samples, whose measured ages exhibit substantial deviations from the expected chronological ordering (presented in Figure~\ref{fig:dag_rocenpail}), thus providing a setting in which to investigate the behavior of the robust projection methods. The luminescence ages used here are part of an ongoing study of the site and are reported with the permission of the excavation team (Soriano et al.). 

\begin{figure}
  \centering
  \includegraphics[width = \linewidth]{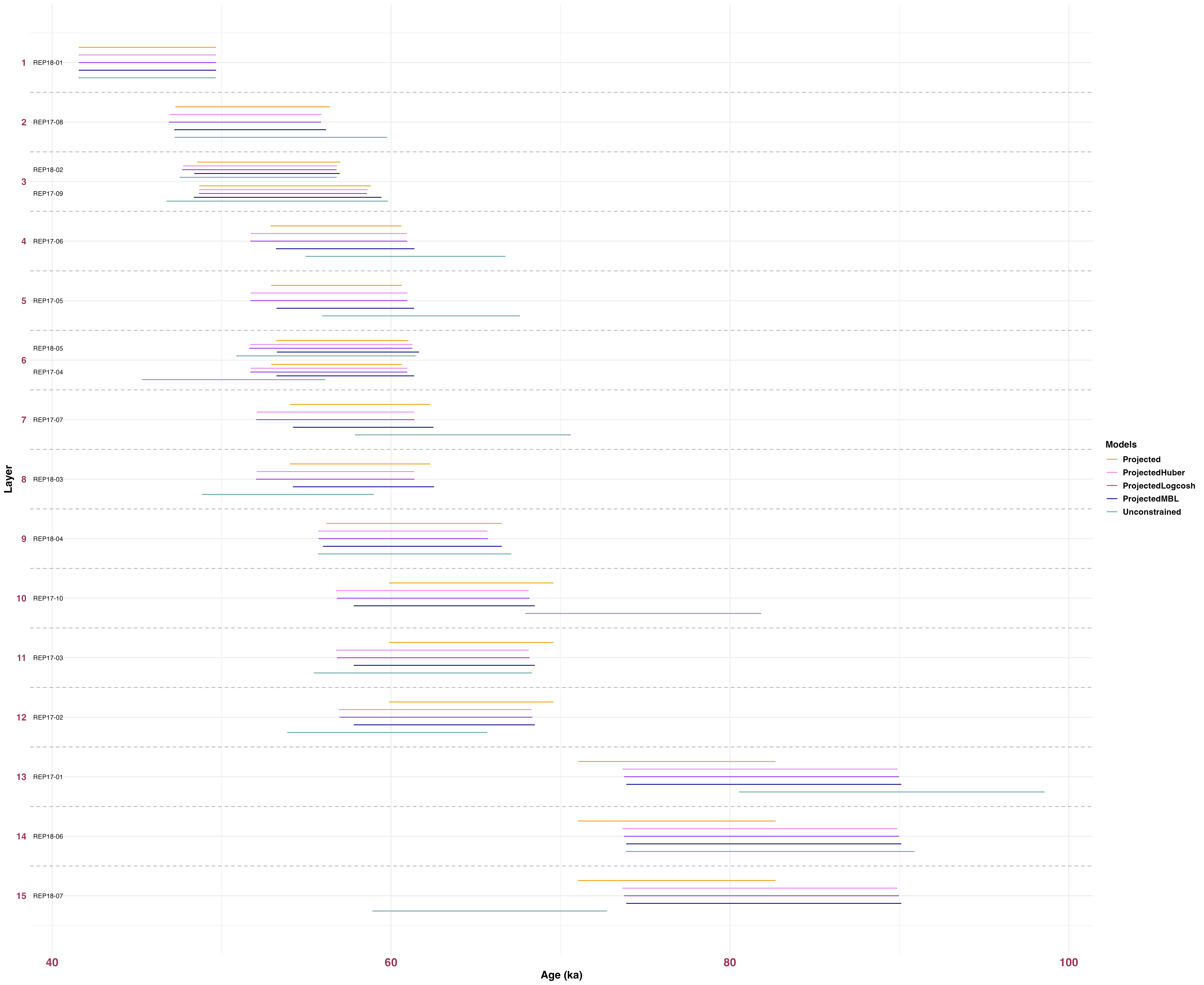}
  \caption{Comparison of the 95\% credible intervals obtained with the Unconstrained, Projected, Projected Log-hyperbolic cosine, Projected MBL for the Roc-en-Pail site.}
  \label{fig:hpd_rocenpail}
\end{figure}

\noindent As can be seen on Figures~\ref{fig:dag_rocenpail} and~\ref{fig:hpd_rocenpail}, the robust models accommodate outlying measurements, in contrast to a mere projection on the constrained space. At layer 10, sample REP17-10 exhibits a marked deviation in the Unconstrained posterior, which leads to an overestimation of the samples at the subsequent layers 11 and 12, seemingly forming a single block. Based on their estimated values, these samples are associated with an extremely cold and dry period.

\noindent A second, more pronounced deviation, and the clearest one with respect to the provided DAG, occurs at layer 15, for sample REP18-08. In this case, the robust models prevent the underestimation of the samples at layers 13 and 14. These samples, based on their estimated values, are associated with the first phase of occupations at the end of the last interglacial cycle.

\noindent The main difference between the non-robust Projected posterior and its robust counterparts is that the latter suggest a larger temporal gap between these two periods and potentially indicating a plausible longer duration of occupation at the end of the last interglacial cycle.      

\subsubsection{Saint-Bélec site}
\begin{figure}
  \centering
  \includegraphics[width = \linewidth]{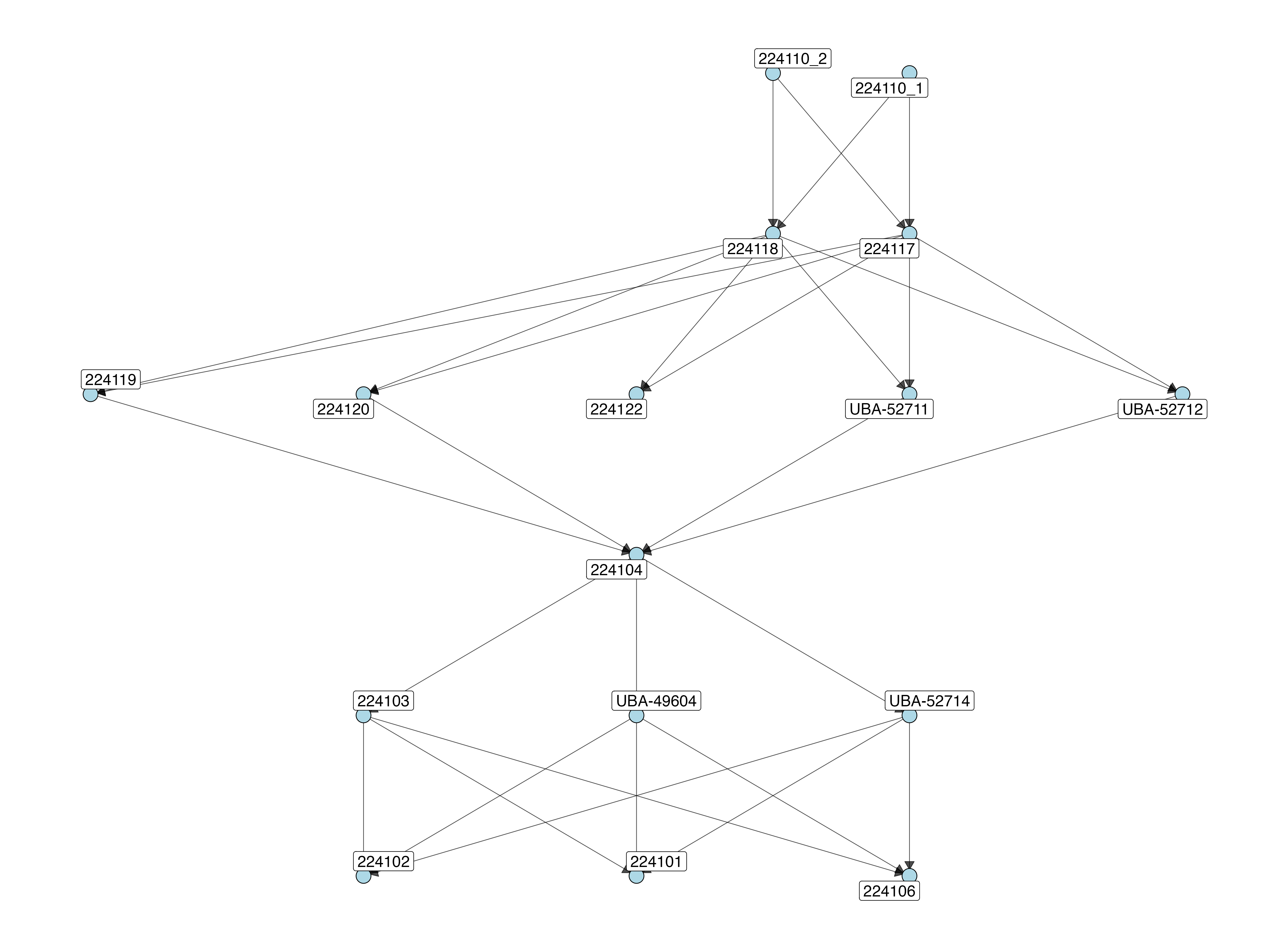}
  \caption{Stratigraphic Structures (DAG representation) of the Saint-Bélec site. }
  \label{fig:net_Stbelec}
\end{figure}

The fourth and last application concerns the Saint-Bélec burial mound, located in Leuhan (Brittany, France). The site is a Bronze Age funerary monument, originally excavated in the early 1900s by Paul du Châtellier and recently investigated through renewed archaeological campaigns \cite{nicolas2021carte}. In this application, we use the available chronological measurements together with the stratigraphic structure represented in Figure~\ref{fig:net_Stbelec}. 

\begin{figure}
  \centering
  \includegraphics[width = \linewidth]{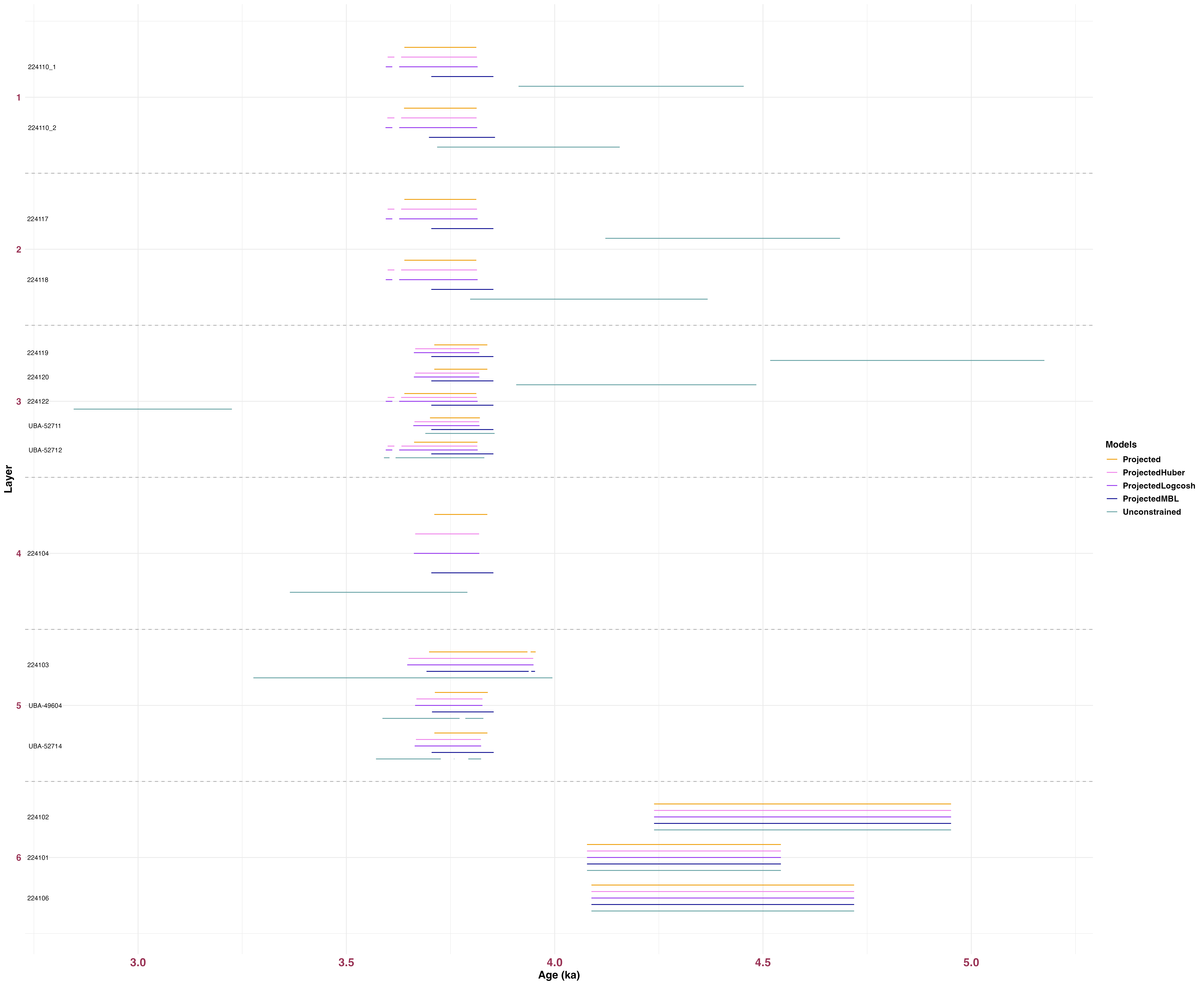}
  \caption{Comparison of the 95\% credible intervals obtained with the Unconstrained, Projected, Projected Log-hyperbolic cosine and Projected MBL for measures at Saint-Bélec site.}
  \label{fig:hpd_Stbelec}
\end{figure}

\noindent Here the dating study combines radiocarbon dating of charcoal with luminescence dating of both sediment and rock surface burial. This combination is interesting because radiocarbon dates human activity on the site, while OSL dating of sediment characterized sediment deposition before site occupation and infilling of the monument; most importantly, rock surface luminescence dating allows estimating the construction age of the burial mound, since the dated event is the last exposure of rocks to light. Since the 14C-dated fireplaces occurred after the construction pit was dug, but before the mound was erected, the combination of all measured ages is expected to significantly refine the chronology of the burial mound, especially since the unconstrained ages (corresponding to calibrated radiocarbon and measured OSL ages, see Figure~\ref{fig:hpd_Stbelec}) are somewhat scattered.

\noindent Two of the tested robust models (ProjectedHuber, ProjectedLogcosh) provide, in this case study, results that are similar to the simple isotonic projection. However, the ProjectedMBL model yields a slightly different estimate for the beginning of the section, as the deviating sample \(224120\) is downweighted.
More importantly, all these models point to a short period of time for the construction period (Note that the top three samples correspond to sterile sediment that was deposited before human occupations). This short period, compared to scattered individual ages, is consistent with our knowledge that the construction of a burial mound can certainly not have spread over decades - let alone centuries. Hence, the tested models provide results that are much more consistent with the archaeological evidence, compared to the set of unconstrained ages.

\section{Conclusion}
Ordering constraints can be encoded either in the prior of a chronological Bayesian model, or in a projection map applied to an unconstrained posterior. In \citet{bouafia2026projectionbasedapproachbayesianage}, the quadratic isotonic regression was used as projection map, which efficiently reduces the duration inflation observed in existent archaeological and geochronological softwares. However, when some observations are incompatible with the imposed order, the quadratic projection pools them with their neighbors, and thereby degrades the estimates of nearby, uncontaminated samples. Robust projection maps are less sensitive to such observations: the isotonic path they return follows the overall stratigraphic structure instead of the outliers. 
\smallskip

\noindent This robustness comes at a theoretical cost: robust maps are not metric projections, so they do not inherit the Wasserstein characterization of the quadratic projected posterior. 
Moreover, the main difficulty is the choice of the robust map's hyperparameter. In fact the block structure of the solution varies with it. To study this dependence, we analyzed the function $\psi$, which counts the observations downweighted by the robust loss. We used the elbow of $\psi$ as a data-driven choice of the hyperparameter. However, the elbow of $\psi$ is a heuristic:  $\psi$ may underestimate contamination when the fit follows the outliers, and it does not directly account for partition changes.

\smallskip
\noindent These robust maps come at little cost when the data contain no outliers: beyond
an explicit threshold, the Huber and MBL maps coincide with the quadratic
projection, the log-hyperbolic cosine map converges
to it at rate $O(\kappa^{-2})$, and their performance in the
uncontaminated scenario is comparable to that of the quadratic projected
posterior. Since the projection acts on each posterior sample, the approach is a
post-processing step that applies to the MCMC output of any unconstrained model, without requiring a new sampler. Our adaptation of the SBM algorithm to the Huber, MBL and log-hyperbolic cosine losses makes our approach computationally inexpensive for arbitrary stratigraphic DAGs.
\smallskip

\noindent This framework should be useful for chronological studies of sites containing
questionable samples: rather than removing them beforehand on subjective grounds,
it keeps them in the analysis and flags them through their spikes.

\section*{Acknowledgements}
This article was supported by the Quina World project (ERC starting grant \#851793 awarded to G.G.). The authors are grateful to J.P. Buylaert for sharing the OSL dataset from Jingbian, China; to S. Soriano for granting us permission to use the OSL dataset from Roc-en-Pail before publication of the site's archaeostratigraphy in a dedicated article; and to T. Freiesleben, K. Thomsen, C. Nicolas and Y. Pailler for letting us use the 14C and OSL data from Saint-Bélec ahead of their publication in an archaeology journal.


	
\clearpage
\bibliography{DOCTORAT}

\end{document}